\documentclass{amsart}
\usepackage{amsfonts,latexsym,amsthm,amssymb,amsmath,amscd,euscript}
\usepackage{framed}
\usepackage[colorlinks=false]{hyperref}
\usepackage{enumitem}
\usepackage{mathrsfs}
\usepackage{enumitem}
\usepackage{amsthm}
\usepackage{MnSymbol}
\usepackage{tikz}
\usetikzlibrary{automata,positioning}
\usepackage{graphicx} 
\usepackage{parskip} 
\usepackage[capitalise]{cleveref} 
\usepackage{csquotes}

\title{Generalizing Cobham's Theorem to $F$-automata}
\author{Laindon Burnett}
\thanks{The arguments in Section 5 from Definition \ref{cellsetup} onward to the proof of Theorem \ref{mainresult3} were proven with assistance from Claude Opus 5.5}

\theoremstyle{plain}
\newtheorem{thm}[subsection]{Theorem}
\newtheorem{lemma}[subsection]{Lemma}
\newtheorem{cor}[subsection]{Corollary}
\newtheorem{prop}[subsection]{Proposition}

\theoremstyle{definition}
\newtheorem{defn}[subsection]{Definition}
\newtheorem{ex}[subsection]{Example}
\newtheorem{rem}[subsection]{Remark}

\newcommand{\BN}{\mathbb N}
\newcommand{\BZ}{\mathbb Z}
\newcommand{\BQ}{\mathbb Q}
\newcommand{\BR}{\mathbb R}
\newcommand{\BC}{\mathbb C}

\newcommand{\Sec}{\mathrm{Sec}}

\newcommand{\Bd}{\mathrm{Bd}}

\usepackage{biblatex}
\begin{document}

\maketitle

\begin{abstract}
We provide two generalizations of Cobham's Theorem from automata theory to the setting of $F$-automatic sequences and sets for a linear map $F: \BZ^d \rightarrow \BZ^d$. We produce as special cases in dimension $2$ a proof of a conjecture of Hansel and Safer distinct from the recent proof by Bustos-Gajardo, Fokkink, and Yassawi, as well as a proof that Cobham's Theorem is satisfied for any imaginary quadratic integers $w$ and $z$ with multiplicatively independent moduli greater than $1$, and with at least one of $w$ and $z$ having no real power. We also provide a classification of which Presburger definable sets are $F$-automatic.
\end{abstract}

\section{Introduction}
We refer the reader to Section $2$ for definitions of terms used in the introduction. 

In 1969, Alan Cobham \cite{originalcobham} proved the following Theorem for automatic sequences: \\

\begin{thm}
Let $k, \ell \geq 2$ be multiplicatively independent natural numbers (there do not exist positive integers $a$ and $b$ such that $k^a = \ell^b$) and let $(a_n)_{n \geq 0}$ be a sequence which is both $k$-automatic and $\ell$-automatic. Then $(a_n)_{n \geq 0}$ is ultimately periodic.
\end{thm}

This is considered to be one of the more important results on automatic sequences. Several authors have attempted to produce Cobham-like results for other mathematical sequences, such as \textit{$k$-regular sequences} \cite{kregular} and \textit{asymptotic automatic sequences} \cite{asymptotic}, and there is a strong version of the theorem in logic \cite{strongcobham}.

One of the difficulties in producing Cobham-like theorems is the difficulty present in the ideas and execution of Cobham's proof. In \cite{Eilenberg}, Samuel Eilenberg says of the theorem:

\begin{displayquote}
``The proof is correct, long, and hard. It is a challenge to find a more reasonable proof of this fine theorem.''
\end{displayquote}

Recently, Krebs \cite{krebs} found such a proof. This proof is much simpler, only a page long, and does not seem to use many properties that require one to use the integers. As a result, one might expect a proof in the style of Krebs to be possible in far more generality.

Hansel and Safer \cite{hanselsafer} attempted to generalise results from automatic sequences to the Gaussian integers, and produced the following conjecture: \\

Let $p$ and $q$ be distinct positive integers. Then a set $S \subseteq \BZ[i]$ is $(-p+i)$ and $(-q+i)$-automatic if and only if there exists a periodic set $P$ such that $S \Delta P$ is finite. 

Here $S \Delta P$ represents the symmetric difference of the sets $S$ and $P$. Krebs claims at the end of his paper that his argument generalises to Gaussian integers, and, more generally, vectors of imaginary quadratic integers. Bustos-Gajardo, Fokkink, and Yassawi \cite{Gaussian26} recently generalized the argument to Gaussian integers and proved Hansel and Safer's conjecture, but there is no known proof for imaginary quadratic integers. We will prove both as special dimension $2$ cases of our first primary result, provided the numbers have multiplicatively independent moduli and at least one number in the pair has no real power. 

Let $\Gamma$ be a finitely generated abelian group and $F: \Gamma \rightarrow \Gamma$ an endomorphism. Intuitively, one can define a number's expansion in ``base $F$" by applying powers of $F$ to the digits, in a similar way to how for $b \in \BN$ a traditional base-$b$ expansion would involve multiplying the digits by powers of $b$. From this, the notion of $F$-automaticity can be derived from the classical notion of automaticity by feeding input to an automaton in base-$F$ rather than base-$b$. Our main result considers automatic sequences in the setting of $F$-automatic sets and answers the conjecture of Hansel and Safer in the affirmative as well as proving the generalisations claimed by Krebs as special cases. It is as follows: \\

\begin{thm}
Let $F, G \in M_{d \times d}(\BZ)$ and let $D_F, D_G \subseteq \BZ^d$ be digit sets. Suppose that the following assumptions hold:

\begin{itemize}
\item $F$ and $G$ commute and are diagonalizable over $\BC$ (equivalently, they are simultaneously diagonalizable).
\item All of the eigenvalues of $F$ over $\BC$ have the same modulus $\lambda > 1$, and all of the eigenvalues of $G$ over $\BC$ have the same modulus $\mu > 1$, with $\lambda$ and $\mu$ being multiplicatively independent.
\item Either there are no non-trivial $F^\ell$-invariant subspaces of $\BQ^d$ for any $\ell \geq 1$, or there are no non-trivial $G^\ell$-invariant subspaces of $\BQ^d$ for any $\ell \geq 1$.
\end{itemize}

Then if $d \geq 2$, a sequence $(f_n)_{n \in \BZ^d}$ is both $(D_F, F)$ and $(D_G, G)$-automatic if and only if it is ultimately periodic with period $H$ for some injective endomorphism $H: \BZ^d \rightarrow \BZ^d$. If instead $d=1$, then a sequence $(f_n)_{n \in \BZ}$ is both $(D_F, F)$ and $(D_G, G)$-automatic if and only if there exists $a, b \in \BZ$ such that the sequence is periodic on $(-\infty, a]$ and periodic on $[b, \infty)$, possibly with different periods.
\end{thm}

The last of these assumptions is not a natural condition, but it is needed in the proof of the theorem, and is enough to prove Hansel and Safer's conjecture as well as Krebs' claim. By altering this assumption to a more natural one, we can use a generalized version of an approach employed by Muchnik in \cite{Muchnik} to prove a slightly weaker statement which is still a generalization of the Cobham-Semenov Theorem. \\

\begin{thm}\label{sec4result}
Let $F, G \in M_{d \times d}(\BZ)$ and let $D_F, D_G \subseteq \BZ^d$ be digit sets. Suppose that the following assumptions hold:

\begin{itemize}
\item $F$ and $G$ commute and are diagonalizable over $\BC$ (equivalently, they are simultaneously diagonalizable).
\item All of the eigenvalues of $F$ over $\BC$ have the same modulus $\lambda > 1$, and all of the eigenvalues of $G$ over $\BC$ have the same modulus $\mu > 1$, with $\lambda$ and $\mu$ being multiplicatively independent.
\item $F$ has $d$ distinct eigenvalues, and $\lambda_i/\lambda_j$ is not a root of unity whenever $i \neq j$.
\end{itemize}

Then a set $A$ that is both $(D_F, F)$ and $(D_G, G)$-automatic is Presburger definable. \\
\end{thm}

Since for integers $k \geq 2$, a Presburger definable set is $k$-automatic, it is natural to ask when a Presburger definable set in $\BZ^d$ has a $(D_F, F)$-automaton which computes it. We establish the following characterization:

\begin{thm}
Let $F$ satisfy the assumptions of Theorem \ref{sec4result} which do not specify a $G$. Then a Presburger definable set $A$ is $(D_F, F)$-automatic for some digit set $D_F$ if and only if $A$ is a finite Boolean combination of sets of the form $v + V$ and $v + (V \cap \{x: \pm \ell(x) \geq 0\})$ with $V$ a finitely generated subgroup of $\BZ^d$, $\mathrm{span}_\BQ V$ $F$-invariant, and with $\ell$ the primitive integral left eigenvector of $F$ with real eigenvalue, unique up to sign, if such an $\ell$ exists.
\end{thm}

In section 2, we outline the preliminaries and definitions needed to speak of $F$-automatic sequences. In section 3, we prove the first of our two results, and in section 4, we prove the second. In section 5, we prove a characterization for which Presburger definable sets are recognizable by a $(D_F, F)$-automaton. Finally, in section 6, we provide relevant examples to which our results apply, and provide a new proof of the conjecture of Hansel and Safer. 

\section{Preliminaries and Notation}
We employ the standard notation for deterministic finite automata (DFAs), namely that of the $5$-tuple $(Q, \Sigma, \delta, q_0, S)$ where $Q$ is the set of states, $\Sigma$ is the input alphabet, $\delta: Q \times \Sigma \rightarrow Q$ is the transition function, $q_0 \in Q$ is the starting state, and $S \subseteq Q$ is the set of accepting states. For a state $q \in Q$ and a word $w \in \Sigma^*$, writing $w = cw'$ with $c \in \Sigma$, we recursively define
\begin{align*}
\delta(q, cw') = \delta(\delta(q, c), w').
\end{align*}
Similarly, a deterministic finite automaton with output (DFAO) is a $6$-tuple $(Q, \Sigma, \delta, q_0, \Delta, \tau)$ with the first four entries being as before, $\Delta$ being a finite output alphabet, and $\tau: Q \rightarrow \Delta$ being the output function. \\

In the traditional setting, we say that a sequence $(a_n)_{n \geq 0}$ is $k$-automatic if there exists a DFAO $M$ reading in inputs from $\Sigma^*$ where $\Sigma = \{0, 1, \ldots, k-1\}$ and which outputs $a_n$ precisely when the input is the base $k$ expansion of $n$. A set $A$ being $k$-automatic can be viewed as the characteristic sequence of $A$, $\chi_A$, being a $k$-automatic sequence, or equivalently, that there is a DFA which accepts precisely the base $k$ expansions of elements from $A$. \\

In generalizing this, let $\Gamma$ be a finitely generated abelian group, let $F: \Gamma \rightarrow \Gamma$ be an injective endomorphism, and let $D \subset \Gamma$ be a finite subset. We take inspiration from \cite{Fsets}, where Bell and Moosa define a base $(D, F)$ expansion of $x$ to be 
\begin{align*}
x = F^m d_m + F^{m-1}d_{m-1} + \cdots + F d_1 + d_0,
\end{align*}
with each $d_j \in D$, and where $F^j$ represents the map $F$ applied $j$ times, with $F^0$ being the identity. Note that multiplication by $k$ can be viewed as a linear map $k: \BZ \rightarrow \BZ$ and is an example of such an $F$, but that unlike base $k$, we need not have uniqueness of representation. Because of this, there is not a canonical digit set $D$ to use either, and so it becomes important to specify which digit set we are working with. We will always assume that $0 \in D$ for any digit set $D$, so that we may pad by zeroes. As a shorthand, if $w = d_md_{m-1}\cdots d_1d_0$, we write this as $x = [w]_F$. \\

By convention, we assume that automata read their inputs from left to right (MSD). It is well known that a right-to-left (LSD) counterpart always exists when the left-to-right version does and vice versa, which we will occasionally employ. 

A natural non-trivial example of such an $F$ is that of Gaussian integers, as seen in the context of Hansel and Safer's work. One can consider the Gaussian integers as vectors in $\BZ^2$, so that multiplication by $a+bi$ is given by the matrix
\begin{align*}
\begin{pmatrix}
a & -b \\
b & a
\end{pmatrix}.
\end{align*}

In \cite{Fsets}, the authors require $D$ to be what is known as an \textit{$F$-spanning set}, which imposes several additional requirements on properties that $D$ can possess. Among these is the requirement that for all $x_1, x_2, x_3, x_4, x_5 \in D$, there exist $t, t' \in D$ such that $x_1 + x_2 + x_3 + x_4 + x_5 = t + Ft'$ (i.e. that the sum of five digits has a representation of length at most $2$ digits). We will not require $D$ to be an $F$-spanning set, as this restricts possible bases for which our results will hold, even in the case of $F$ simply being multiplication by a non-integer. \\

\begin{ex}
In \cite{-n+i}, the authors establish that for $p$ a positive integer, every element of $\BZ[i]$ has a unique base $(-p+i)$ expansion with digits from $\{0, 1, \ldots, p^2\}$. Letting $p = 2$, we have that 
\begin{align*}
4 + 4 + 4 + 4 + 0 = (-2+i)^5 + 3(-2+i)^4 + 4(-2+i)^3 + 4(-2+i)^2 + 3(-2+i) + 1,
\end{align*}
which is certainly more than two digits long.
\end{ex}

The above aside, our definitions of $F$-automatic sequences and sets will be very similar to \cite{Fsets}.  Since there is no one canonical way of ordering elements in our generalized setting, it is easier to view a sequence $(f_n)_{n \in \Gamma}$ taking values in a set $Y$ as a function $f: \Gamma \rightarrow Y$. We will at times write $f_n$ to mean $f(n)$. \\

\begin{defn}
Let $(D, F)$ be a complete numeration system, so that every $x \in \Gamma$ has at least one base $(D, F)$-representation. A sequence $f: \Gamma \rightarrow Y$ is $(D, F)$-automatic if there is a DFAO $M$ such that for all $x \in \Gamma$, and all base $(D, F)$ expansions $x = [w]_F$, $M$ on input $w$ outputs $f(x)$. Similarly, a set $A$ is $(D, F)$-automatic if there is a DFA $M$ such that for  such that for all $x \in \Gamma$, and all base $(D, F)$ expansions $x = [w]_F$, $M$ accepts input $w$ if and only if $x \in A$.  
\end{defn}

We will show in Lemma \ref{digitsetlemma} and Lemma \ref{expanddigitlemma} that a digit set $D$ can always be enlarged sufficiently to some digit set $D'$ to form a complete numeration system $(D', F)$, so we do not lose anything from this assumption.

We must also make sense of what it means to be periodic or ultimately periodic in our generalized setting. The period must also be generalized: it is no longer multiplication by a number, but also an injective endomorphism. \\

\begin{defn}
Let $H: \Gamma \rightarrow \Gamma$ be an endomorphism of finitely generated abelian groups. We say that a sequence $(f_n)_{n \in \Gamma}$ is periodic with period $H$ if $f_{n + Hx} = f_n$ for all $n, x \in \Gamma$, and is ultimately periodic if it differs from a periodic sequence on finitely many terms. A set $P$ is periodic with period $H$ if $n \in P \iff n + Hx \in P$ for all $n, x \in \Gamma$. A set $X$ is ultimately periodic if there is a periodic set $P$ such that $X \Delta P$ is finite. 
\end{defn}

\section{Generalization of Krebs' Proof}
\begin{thm}\label{mainresult1}
Let $F, G \in M_{d \times d}(\BZ)$ and let $D_F, D_G \subseteq \BZ^d$ be digit sets. Suppose that the following assumptions hold:

\begin{itemize}
\item $F$ and $G$ commute and are diagonalizable over $\BC$ (equivalently, they are simultaneously diagonalizable).
\item All of the eigenvalues of $F$ over $\BC$ have the same modulus $\lambda > 1$, and all of the eigenvalues of $G$ over $\BC$ have the same modulus $\mu > 1$, with $\lambda$ and $\mu$ being multiplicatively independent.
\item Either there are no non-trivial $F^\ell$-invariant subspaces of $\BQ^d$ for any $\ell \geq 1$, or there are no non-trivial $G^\ell$-invariant subspaces of $\BQ^d$ for any $\ell \geq 1$.
\end{itemize}

Then if $d \geq 2$, a sequence $(f_n)_{n \in \BZ^d}$ is both $(D_F, F)$ and $(D_G, G)$-automatic if and only if it is ultimately periodic with period $H$ for some injective endomorphism $H: \BZ^d \rightarrow \BZ^d$. If instead $d=1$, then a sequence $(f_n)_{n \in \BZ}$ is both $(D_F, F)$ and $(D_G, G)$-automatic if and only if there exists $a, b \in \BZ$ such that the sequence is periodic on $(-\infty, a]$ and periodic on $[b, \infty)$, possibly with different periods.
\end{thm}

It is not immediately clear why the last of these assumptions is necessary. In \cite{bosmafokkinkkrebs}, the authors show that there exists a complete and unambiguous numeration system for any $z \in \BZ[i]$ with $|z| \geq \sqrt{5}$ using the digit set
\begin{align*}
D = \left\{d \in \BZ[i]: -\frac{1}{2} \leq \mathrm{Re}(d/z) < \frac{1}{2}, -\frac{1}{2} \leq \mathrm{Im}(d/z) < \frac{1}{2}\right\}.
\end{align*}
From this, they show that $\BZ$ as a subset of $\BZ[i]$ is $z$-automatic with digit set $D$ if and only if $z^j \in \BN$ for some $j \in \BN$. In particular, if we take two multiplicatively independent integers such as $3$ and $5$ and view their multiplications over $\BZ[i]$ as maps from $\BZ^2$ to $\BZ^2$, then $\BZ$ is $3$ and $5$-automatic with these respective digit sets. However, $\BZ$ is not an ultimately periodic set with respect to any injective endomorphism on $\BZ^2$, since any such endomorphism would need to send every Gaussian integer to an integer, meaning it would be rank one. As matrices in $M_{2 \times 2}(\BZ)$, both $3$ and $5$ satisfy the first two assumptions of the theorem, but violate the third, since they send $\BQ \times \{0\}$ to $\BQ \times \{0\}$. \\

We will present a second result later which allows for invariant subspaces, with an additional third assumption instead. \\

Most of our arguments will take advantage of simultaneous diagonalizability of $F$ and $G$, and so we will make use of an eigenvector basis. We require the following definition. \\

\begin{defn}
Let $V = \{v_1, \ldots, v_d\}$ be the eigenvector basis corresponding to the simultaneous diagonalization of $F$ and $G$. Given $x = c_1v_1 + \cdots + c_dv_d \in \BC^d$ and $A \in M_{d \times d}(\BC)$, define $\|x\|_V = \max\limits_{1 \leq j \leq d} |c_j|$ and $\|A\|_V = \sup\{\|Ax\|_V: \|x\|_V = 1\}$. We will refer to this norm as the $V$-norm. \\
\end{defn}

\begin{lemma}\label{closepowerslemma}
Let $\epsilon > 0$, and let $F, G \in M_{d \times d}(\BZ)$ be simultaneously diagonalizable over $\BC$, with $F$ having all eigenvalues of modulus $\lambda$ and $G$ having all eigenvalues of modulus $\mu$ for some multiplicatively independent $\lambda, \mu > 1$. Then for any $\eta > 0$, there exist infinitely many pairs of constants $(m, n) \in \BN^2$ such that
\begin{align*}
\|F^m - G^n\|_V \leq \epsilon\min\{\lambda^m, \mu^n\}
\end{align*}
and $\max\{\lambda^m, \mu^n\} \leq (1+\eta)\min\{\lambda^m, \mu^n\}$.
\end{lemma}

\begin{proof}
Write the eigenvalues of $F$ as $\lambda e^{i\theta_1}, \ldots, \lambda e^{i\theta_d}$ for some $\theta_1, \ldots, \theta_d \in [0, 2\pi)$ and the eigenvalues of $G$ as $\mu e^{i\psi_1}, \ldots \mu e^{i\psi_d}$ for some $\psi_1, \ldots, \psi_d \in [0, 2\pi)$. Note that for any integer $q \geq 1$, we may write
\begin{align*}
|\lambda^{qm}e^{iqm\theta_j} - \mu^{qn}e^{iqn\psi_j}| &\leq |(\lambda^{qm} - \mu^{qn})e^{iqm\theta_j}| + |\mu^{qn}(e^{iqm\theta_j} - e^{iqn\psi_j})| \\
&= |\lambda^{qm} - \mu^{qn}||e^{iqm\theta_j}| + |\mu^{qn}||e^{iqm\theta_j} - e^{iqn\psi_j}| \\
&= |\lambda^{qm} - \mu^{qn}| + \mu^{qn}|e^{iqm\theta_j} - e^{iqn\psi_j}|.
\end{align*}
Similarly, we have that
\begin{align*}
|\lambda^{qm}e^{iqm\theta_j} - \mu^{qn}e^{iqn\psi_j}| \leq |\lambda^{qm} - \mu^{qn}| + \lambda^{qm}|e^{iqm\theta_j} - e^{iqn\psi_j}|,
\end{align*}
and so in particular,
\begin{align}
|\lambda^{qm}e^{iqm\theta_j} - \mu^{qn}e^{iqn\psi_j}| \leq |\lambda^{qm} - \mu^{qn}| + \min\{\lambda^{qm}, \mu^{qn}\}|e^{iqm\theta_j} - e^{iqn\psi_j}|. \label{multindeptarget}
\end{align}
We now bound both $|\lambda^{qm} - \mu^{qn}|$ and $|e^{iqm\theta_j} - e^{iqn\psi_j}|$. \\

Since $\lambda$ and $\mu$ are multiplicatively independent, $\frac{\log \lambda}{\log \mu}$ is irrational. By Dirichlet's Approximation Theorem, there exist infinitely many positive integers $m$ and $n$ such that
\begin{align*}
0 < \left| \frac{\log \lambda}{\log \mu} - \frac{n}{m}\right| < \frac{1}{m^2} &\implies 0 < |m\log \lambda - n\log \mu| < \frac{1}{m}\log \mu \\
&\implies 1 < \exp(|m \log \lambda - n\log \mu|) < \exp\left(\frac{1}{m}\log \mu\right).
\end{align*}
If $m \log \lambda - n \log \mu > 0$, or equivalently $\lambda^m > \mu^n$, the above becomes
\begin{align*}
1 < \frac{\lambda^m}{\mu^n} < \mu^{1/m},
\end{align*}
and Bernoulli's inequality gives that $\mu^{1/m} < 1 + \frac{\mu - 1}{m}$, so that
\begin{align*}
\left|\frac{\lambda^m}{\mu^n} - 1\right| < \frac{\mu - 1}{m}.
\end{align*}
Multiplying through by $\mu^n$ gives that 
\begin{align*}
|\lambda^m - \mu^n| < \frac{\mu - 1}{m}\mu^n,
\end{align*}
and since $\lambda^m > \mu^n$, we obtain that
\begin{align}
|\lambda^m - \mu^n| < \frac{\mu - 1}{m}\min\{\lambda^m, \mu^n\}. \label{multindepineq1}
\end{align}
By a similar argument, if $\mu^n > \lambda^m$ we also obtain the above inequality. Regardless of which case we are in, we also have that
\begin{align}
1 < \frac{\max\{\lambda^m, \mu^n\}}{\min\{\lambda^m, \mu^n\}} < 1 + \frac{\mu - 1}{m}. \label{multindepineq2}
\end{align}
Next we will bound $|e^{iqm\theta_j} - e^{iqn\psi_j}|$. Fixing $m$ and $n$ from the above, let $u_j = m\theta_j \mod 2\pi$ and $v_j = n\psi_j \mod 2\pi$, and note that the angle between $e^{iqm\theta_j}$ and $e^{iqn\psi_j}$ is $|qu_j - qv_j| \mod 2\pi$, which is also the length of the arc on the unit circle between these two numbers. This is an overestimate for their distance. By a multidimensional analogue of Dirichlet's Approximation Theorem, for any $N \in \BN$, we may find $p_1, \ldots, p_d, p_1', \ldots p_d' \in \BZ$ and $q' \in \BZ$ with $1 \leq q' \leq N^{2d}$ such that
\begin{align*}
& \left| \frac{u_j}{2\pi} - \frac{p_j}{q'}\right| < \frac{1}{q'N} \text{ and } \left|\frac{v_j}{2\pi} - \frac{p_i'}{q'}\right| < \frac{1}{q'N} \text{ for } 1 \leq j \leq d \\
\iff & |q'u_j - p_j2\pi| < \frac{2\pi}{N} \text{ and } |q'v_j - p_j'2\pi| < \frac{2\pi}{N} \text{ for } 1 \leq j \leq d,
\end{align*}
where $2\pi/N < \epsilon/4$ for $N$ sufficiently large. Note that while the value of $q'$ depends on our choice of $m$ and $n$, we know that $q' < N^{2d}$, and the choice of $N$ is independent of $m$ and $n$. We will take the arbitrary $q$ from the previous argument to be equal to the value of $q'$ which we have just found. We obtain that
\begin{align*}
|e^{iq'm\theta_j} - e^{iq'n\psi_j}| &= |e^{iq'u_j} - e^{iq'v_j}| \\
&\leq |q'u_j - q'v_j| \mod 2\pi \\
&= |q'u_j - p_j2\pi + p_j'2\pi - q'v_j| \mod 2\pi \\
&\leq |q'u_j - p_j2\pi| + |q'v_j - p_j'2\pi| \\
& < \epsilon/4 + \epsilon/4 \\
&= \epsilon/2. \\
\end{align*}

Focusing on the first piece once again and taking $q$ to be the value of $q'$ from above, we have that
\begin{align*}
|\lambda^{q'm} - \mu^{q'n}| &\leq |\lambda^m - \mu^n||(\lambda^m)^{q'-1} + (\lambda^m)^{q'-2}\mu^n + \cdots + (\mu^n)^{q'-1}| \\
&\leq |\lambda^m - \mu^n|q'(\max\{\lambda^m, \mu^n\})^{q'-1} \\
&\leq \frac{\mu - 1}{m}\min\{\lambda^m, \mu^n\}q'(\max\{\lambda^m, \mu^n\})^{q'-1} \text{ by } \eqref{multindepineq1} \\
&< (\mu - 1)\frac{q'}{m}(\max\{\lambda^m, \mu^n\})^{q'} \\
&= (\mu - 1)\frac{q'}{m}(\min\{\lambda^m, \mu^n\})^{q'}\left(\frac{\max\{\lambda^m, \mu^n\}}{\min\{\lambda^m, \mu^n\}}\right)^{q'} \\
&< (\mu - 1)\frac{q'}{m}\min\{\lambda^{q'm}, \mu^{q'
n}\}\left(1 + \frac{\mu - 1}{m}\right)^{q'} \text{ by } \eqref{multindepineq2} \\
&\leq (\mu - 1)\frac{N^{2d}}{m}\min\{\lambda^{q'm}, \mu^{q'n}\}\left(1 + \frac{\mu - 1}{m}\right)^{N^{2d}} \text{ since } q' \leq N^{2d} \\
&\leq \frac{\epsilon}{2}\min\{\lambda^{q'm}, \mu^{q'n}\} \text{ for large $m$},
\end{align*} 
since as $m$ grows large, $\frac{N^{2d}(\mu - 1)}{m}(1 + \frac{\mu - 1}{m})^{N^{2d}}$ approaches $0$, as $N$ is independent of our choice of $m$. \\

Combining both pieces with \eqref{multindeptarget}, taking our choice of $q$ to be the $q'$ which made one of our inequalities hold, we obtain that
\begin{align*}
|\lambda^{q'm}e^{iq'm\theta_j} - \mu^{q'n}e^{iq'n\psi_j}| &\leq |\lambda^{q'm} - \mu^{q'n}| + \min\{\lambda^{q'm},  \mu^{q'n}\}|e^{iq'm\theta_j} - e^{iq'n\psi_j}| \\
&< \frac{\epsilon}{2}\min\{\lambda^{q'm}, \mu^{q'n}\} + \frac{\epsilon}{2}\min\{\lambda^{q'm}, \mu^{q'n}\} \\
&= \epsilon\min\{\lambda^{q'm}, \mu^{q'n}\}.
\end{align*}
Take $q'm$ and $q'n$ to be $m$ and $n$ in the statement of the theorem. Since Dirichlet's Approximation Theorem gave us infinitely many valid $m$ and $n$ in the first place, and since we only need to take $m$ sufficiently large, this gives infinitely many valid $q'm$ and $q'n$. If $x \in \BC^d$ with $\|x\|_V = 1$, then $x = \sum_{j=1}^d c_jv_j$ with $|c_j| \leq 1$ for $1 \leq j \leq d$ and equality in at least one index. We have that
\begin{align*}
(F^{q'm} - G^{q'n})x &= \sum\limits_{j=1}^d c_j(F^{q'm} - G^{q'n})v_j \\
&= \sum\limits_{j=1}^d c_j(\lambda^{q'm}e^{iq'm\theta_j} - \mu^{q'n}e^{iq'n\psi_j})v_j,
\end{align*}
so that
\begin{align*}
\|(F^{q'm} - G^{q'n})x\|_V &= \max\limits_{1 \leq j \leq d} |c_j(\lambda^{q'm}e^{iq'm\theta_j} - \mu^{q'n}e^{iq'n\psi_j})| \\
&< \max\limits_{1 \leq j \leq d} (|c_j|\epsilon\max\{\lambda^{q'm}, \mu^{q'n}\}) \\
&= \epsilon\max\{\lambda^{q'm}, \mu^{q'n}\}\max\limits_{1 \leq j \leq d}|c_j| \\
&= \epsilon\max\{\lambda^{q'm}, \mu^{q'n}\}.
\end{align*}
Then $\|F^{q'm} - G^{q'n}\|_V = \sup\{\|(F^{q'm} - G^{q'n})x\|_V: \|x\|_V = 1\} \leq \epsilon\max\{\lambda^{q'm}, \mu^{q'n}\}$.

But by \eqref{multindepineq2}, we have that 
\begin{align*}
\max\{\lambda^{q'm}, \mu^{q'n}\} < \left(1 + \frac{\mu - 1}{m}\right)^{q'}\min\{\lambda^{q'm}, \mu^{q'n}\} < \left(1 + \frac{\mu - 1}{m}\right)^{N^{2d}}\min\{\lambda^{q'm}, \mu^{q'n}\},
\end{align*}
and so by absorbing the constant inside of $\epsilon$, we attain the desired result. Also, for any $\eta > 0$, taking $m$ sufficiently large gives that $(1+\frac{\mu - 1}{m})^{N^{2d}} < (1+\eta)$.
\end{proof}

\begin{defn}
For $C \in \BR$, let $B_V[C] = \{c_1v_1 + \cdots + c_dv_d: \max\limits_{1 \leq j \leq d}|c_j| \leq C\}$, with the $c_j \in \BC$, which is the ball of radius $C$ around the origin with respect to the $V$-norm. \\
\end{defn}

\begin{lemma}\label{ballimage}
For every $C \in \BR$, we have that $F(B_V[C]) = B_V[\lambda C]$.
\end{lemma}

\begin{proof}
We have that
\begin{align*}
\sum\limits_{j=1}^d c_jv_j \in B_V[C] &\iff |c_j| \leq C \text{ for } 1 \leq j \leq d \\
&\iff |\lambda_jc_j| = \lambda|c_j| \leq \lambda C \text{ for } 1 \leq j \leq d \\
&\iff F\left(\sum\limits_{j=1}^d c_jv_j\right) = \sum\limits_{j=1}^d \lambda_jc_jv_j \in B_V[\lambda C].
\end{align*}
\end{proof}

We wish to establish bounds on how long the minimum length of a digit expansion for a given vector can be. Since we will use this idea in Section 4 on more general linear maps $F$, we prove this for a more general $F$ here.

\begin{lemma}\label{digitsetlemma}
Let $\Gamma$ be a free abelian group of finite rank, and let $F: \Gamma \rightarrow \Gamma$ have all eigenvalues with modulus $\lambda > 1$. Given any digit set $D \subset \Gamma$, there exists a digit set $D' \supset D$ and a divergent sequence of constants $(C_n)_{n \geq 1}$ such that if $x \in \Gamma$ lies within $B_V[C_n]$, then $x$ has a base-$F$ expansion with digits from $D'$ with length at most $n$. In particular, if $F$ is diagonalizable, taking $\|.\|_\infty$ with respect to a fixed $\BZ$-basis of $\Gamma$ and $\|.\|_V$ with respect to any eigenbasis of $F \otimes \BC$, if we let
\begin{align*}
L = \max\limits_{x \in \Gamma: \|x\|_\infty \leq 1} \|x\|_V,
\end{align*}
we may choose such a sequence with $C_n > L\lambda^n$ for all $n \geq 2$.
\end{lemma}

\begin{proof}
Since $\Gamma$ is a finitely generated abelian group and $F: \Gamma \rightarrow \Gamma$ is an injective endomorphism (since no eigenvalues are $0$), then $F(\Gamma)$ is of finite index in $\Gamma$, and so $\Gamma/F(\Gamma)$ is finite. For each coset, choose a representative $x$, and let $N_1$ be the maximum of $\|x\|_V$ across all representatives. Let $N_2 = \max\limits_{x \in D} \|x\|_V$, and take $N$ larger than $\max\{N_1, N_2\}$. Define $D' = B_V[N] \cap \Gamma$. Since $N > N_2$, $D' \supset D$. We show that this choice of $D'$ has the desired properties. Write
\begin{align*}
C_n = N\lambda^{n-1} - N_1\frac{\lambda^{n-1} - 1}{\lambda - 1},
\end{align*}
so that in particular $C_1 = N$. We proceed by induction, with the base case clear since all of $B_V[N]$ can be written as a single digit from $D'$ by definition. Suppose the assertion holds for some $n$. Let $y \in B_V[\lambda C_n - N_1]$. Then $y$ belongs to some coset of $\Gamma/F(\Gamma)$, and so there exists $x \in B_V[N_1]$ such that $y + x \in F(\Gamma)$. Applying the triangle inequality over all $d$ of the scalars,
\begin{align*}
y + x \in B_V[\lambda C_n - N_1 + N_1] = B_V[\lambda C_n] = F(B_V[C_n]),
\end{align*}
with the last equality following from Lemma \ref{ballimage}. By the inductive hypothesis, we may write
\begin{align*}
y + x = F\left(\sum\limits_{j=0}^{n-1} F^jd_j\right) = \sum\limits_{j=1}^n F^jd_{j-1}
\end{align*}
for some $d_0, \ldots, d_{n-1} \in D'$. Since $x \in B_V[N_1]$, $-x \in B_V[N_1]$ also, so that
\begin{align*}
y = \left(\sum\limits_{j=1}^n F^jd_{j-1}\right) - x,
\end{align*}
while we have that
\begin{align*}
\lambda C_n - N_1 &= \lambda\left(N\lambda^{n-1} - N_1\frac{\lambda^{n-1} - 1}{\lambda - 1}\right) - N_1 \\
&= N\lambda^n - N_1\frac{\lambda^n - 1}{\lambda - 1} \\
&= C_{n+1},
\end{align*}
hence the assertion holds for $n+1$ as well. \\

It remains to show that by taking $N$ sufficiently large, we have $L\lambda^n < C_n$ for all $n$. We have that
\begin{align*}
L\lambda^n < C_n &\iff L\lambda^n < N\lambda^{n-1} - N_1\frac{\lambda^{n-1} - 1}{\lambda - 1} \\
&\iff 0 < (N/\lambda - L)\lambda^n - N_1\frac{\lambda^{n-1} - 1}{\lambda - 1} \\
&\iff 0 < (N/\lambda - L)\lambda^{n+1} - (N/\lambda - L)\lambda^n - N_1\lambda^{n-1} + N_1 \\
&\iff 0 < (N/\lambda - L)\lambda^{n+1} - (N/\lambda + N_1/\lambda - L)\lambda^n + N_1
\end{align*}
since $\lambda > 1$. Since $N_1 > 0$, it suffices to have
\begin{align*}
&(N/\lambda - L)\lambda^{n+1} - (N/\lambda + N_1/\lambda - L)\lambda^n > 0 \\
\iff &(N/\lambda - L)\lambda - (N/\lambda + N_1/\lambda - L) > 0 \\
\iff & \lambda > \frac{N/\lambda + N_1/\lambda - L}{N/\lambda - L} \text{ provided that } N/\lambda > L \\ 
\iff & \lambda > 1 + \frac{N_1/\lambda}{N/\lambda - L},
\end{align*}
where the latter term becomes arbitrarily small for $N$ sufficiently large, since $N_1$ is a constant. As $\lambda > 1$, the result follows. Since $L\lambda^n$ diverges as $n \to \infty$, so does $C_n$. 
\end{proof}

With this lemma, we now have a way of guaranteeing a digit set $D$ for which every integer vector has a finite $(D, F)$-expansion, and for which we can bound the length of these expansions. The problem that arises is that our DFA for $F$ may not have a nice digit set like the one constructed in Lemma \ref{digitsetlemma}. The next Lemma will show that given any DFA, we can always expand the digit set in any way and still find a working DFA. In particular, we can expand to the nice digit set that guarantees us finite expansions. \\

\begin{lemma}\label{expanddigitlemma}
Let $M$ be a DFA in base $F$ with digit set $D \subset \BZ^d$ which accepts a set $A \subseteq \BZ^d$. Let $D' \supset D$. Then there exists a DFA $M'$ in base $F$ with digit set $D'$ which accepts $A$. 
\end{lemma}

\begin{proof}
Define $L_{D', D} = \{(v, w) \in (D' \times D)^*: [v]_F = [w]_F\}$. We first prove that this language is regular. We will construct an automaton reading from right to left. Let $(v_1, w_1)\cdots (v_n, w_n)$ be our input. Begin in a state labelled $S_0$. We wish to determine whether $[v_1 \cdots v_n]_F = [w_1 \cdots w_n]_F$, or equivalently, whether
\begin{align*}
& F^{n-1}v_1 + \cdots + Fv_{n-1} + v_n = F^{n-1}w_1 + \cdots + Fw_{n-1} + w_n \\
\iff & F^{n-1}(v_1 - w_1) + \cdots + F(v_{n-1} - w_{n-1}) + (v_n - w_n) = 0 \\
\iff & F(F^{n-2}(v_1 - w_1) + \cdots + (v_{n-1} - w_{n-1})) = w_n - v_n.
\end{align*}
The left side of the last line is of the form $F(u)$ for some $u \in \BZ^d$, so read $(v_n, w_n)$ and compute $u = F^{-1}(w_n - v_n) \in \BQ^d$. If this is not in $\BZ^d$, then the equality is false, and so we transition to a rejecting state $S_{\mathrm{rej}}$. Otherwise, we can apply $F^{-1}$ to both sides, since $F$ is injective, and transition to state $S_u$. 

Now suppose that we are in state $S_u$ with input $(v_1, w_1) \cdots (v_i, w_i)$ left to be read. We have that
\begin{align*}
& F^{i-1}(v_1 - w_1) + \cdots + F(v_{i-1} - w_{i-1}) + (v_i - w_i) = u \\
\iff &  F(F^{i-2}(v_1 - w_1) + \cdots + (v_{i-1} - w_{i-1})) = u + w_i - v_i.
\end{align*}
Read $(v_i, w_i)$ and compute $u' = F^{-1}(u + w_i - v_i)$. If this is not in $\BZ^d$, then again the equality is false, so transition to $S_{\mathrm{rej}}$, otherwise apply $F^{-1}$ to both sides and transition to $S_{u'}$. We should accept after having read all input exactly when we are in state $S_0$, as this means that $F^{-1}(u + w_1 - v_1) = 0$, and so $u = v_1 - w_1$. \\

It remains to show that there are finitely many states $S_u$ which need to be constructed. Choose $C \in \BR$ large enough so that $D - D' \subseteq B_V[C]$. Since $F$ has all eigenvalues of modulus $\lambda > 1$, $F^{-1}$ has all eigenvalues of modulus $1/\lambda < 1$. By Lemma \ref{ballimage}, if $w_n \in D$ and $v_n \in D'$, then $F^{-1}(w_n - v_n) \in B_V[C/\lambda]$. Suppose that after reading $m$ digits of input, we are in state $u \in B_V\left[C\sum\limits_{j=1}^m \lambda^{-j}\right]$. If $w_i \in D$ and $v_i \in D'$, then by triangle inequality \begin{align*}
u + w_i - v_i \in B_V\left[C + C\sum\limits_{j=1}^m \lambda^{-j}\right] = B_V\left[C\sum\limits_{j=0}^m \lambda^{-j}\right],
\end{align*}
and so by Lemma \ref{ballimage},
\begin{align*}
F^{-1}(u + w_i - v_i) \in B_V\left[C/\lambda\sum\limits_{j=0}^m \lambda^{-j}\right] = B_V\left[C\sum\limits_{j=1}^{m+1} \lambda^{-j}\right].
\end{align*}
In particular, for any $m$ we have that $\sum\limits_{j=1}^m \lambda^{-j} < \sum\limits_{j=1}^\infty \lambda^{-j} = \frac{\lambda^{-1}}{1 - \lambda^{-1}}$ since $\lambda^{-1} < 1$. Then any state $S_u$ we ever transition to has $u \in B_V[C\lambda^{-1}/(1 - \lambda^{-1})] \cap \BZ^d$, and so there are finitely many such states. Hence $L_{D', D}$ is a regular language. \\

Next, we consider the projection homomorphisms $\pi_1: (D' \times D)^* \rightarrow (D')^*$ and $\pi_2: (D' \times D)^* \rightarrow D^*$ which project onto the first and second coordinates of words respectively. Let $A_F$ denote the set of words $w \in D^*$ such that $[w]_F \in A$, so that $A_F$ is recognized by $M$. Recall that the image of a regular language under homomorphisms and inverse homomorphisms remains regular, and that the left quotient of two regular languages is regular. Then we have that
\begin{align*}
&A_F \text{ is regular} \\
&\implies \pi_2^{-1}(A_F) = \{(v, w) \in (D' \times D)^*: [w]_F \in A\} \text{ is regular} \\
&\implies L_{D', D} \cap \pi_2^{-1}(A_F) = \{(v, w) \in (D' \times D)^*: [v]_F = [w]_F \in A\} \text{ is regular} \\
&\implies \pi_1(L_{D', D} \cap \pi_2^{-1}(A_F)) \text{ is regular} \\
&\implies 0^*\backslash \pi_1(L_{D', D} \cap \pi_2^{-1}(A_F)) \text{ is regular},
\end{align*}
where 
\begin{align*}
0^*\backslash \pi_1(L_{D', D} \cap \pi_2^{-1}(A_F)) &= \{0^kv: v \in \pi_1(L_{D', D} \cap \pi_2^{-1}(A_F)) \text{ for some } k \geq 0\} \\
&= \{v \in (D')^*: [v]_F \in A\}.
\end{align*}
Taking the left quotient handles the case where every $D$-expansion of $[v]_F$ has length strictly above $|v|$: we can pad $v$ by zeroes to achieve $[0^kv]_F = [v]_F$ of an equal length to some $D$-expansion. Hence there exists a DFA $M'$ in base $(D', F)$ which accepts $A$. 
\end{proof}

\begin{cor}\label{expanddigitdfao}
Suppose that $(a_n)_{n \in \BZ^d}$ is an $F$-automatic sequence with digit set $D \subset \BZ^d$, and that $D' \supset D$. Then this sequence is $F$-automatic with digit set $D'$ as well. 
\end{cor}

\begin{proof}
Let $M$ be a DFAO with input in base $(D, F)$ and which outputs $(a_n)_{n \in \BZ^d}$. Let $\Delta = \{d_1, \ldots, d_r\}$ be the set of output values for $M$, that is, the set of possible values our sequence takes on. For $1 \leq j \leq r$, let $L_{d_j}$ be the set of inputs from $D^*$ to $M$ which produce an output of $d_j$. By Theorem 4.3.1 of \cite{allouche2003automatic}, each $L_{d_j}$ is regular. Let $M_{d_j}$ be a DFA for $L_{d_j}$. By Lemma \ref{expanddigitlemma}, there exists a DFA $M_{d_j}'$ in base $F$ with digit set $D'$ accepting $L_{d_j}$. This means that the set of inputs from $(D')^*$ which $M_{d_j}'$ accepts is a regular language, say $L_{d_j}'$. We have that $v \in L_{d_j}'$ if and only if there exists $w \in L_{d_j}$ such that $[v]_F = [w]_F$ and $M$ outputs $d_j$ on input $w$. Then the $L_{d_j}'$ are disjoint, since otherwise there are $w, w' \in D^*$ such that $[v]_F = [w]_F = [w']_F$ but $M$ outputs different values on inputs $w$ and $w'$, meaning $M$ is not well-defined, a contradiction. Since $M$ computes $(a_n)_{n \in \BZ^d}$ and $a_n = d_j$ for some $n \in \BZ^d$, there is some $w \in D^*$ such that $[w]_F = n$ and $M$ outputs $d_j$ with input $w$, so that every $v \in (D')^*$ belongs to $L_{d_j}'$ for some $j$. Hence the $L_{d_j}'$ form a partition of $(D')^*$. By Theorem 4.3.2 of \cite{allouche2003automatic}, there is an automaton $M'$ which outputs $d_j$ on $L_{d_j}'$. By design, this means that whenever $[v]_F = [w]_F = n$ for some $v \in (D')^*$ and $w \in D^*$, $M'$ on input $v$ and $M$ on input $w$ both produce the same output, namely $a_n$. Hence $M'$ computes $(a_n)_{n \in \BZ^d}$.
\end{proof}

With the above lemma established, we have enough to prove the backward direction of Theorem \ref{mainresult1}. 

\begin{proof}[Proof of Theorem \ref{mainresult1}, backward direction]

We first deal with the case $d \geq 2$. Let $H: \BZ^d \rightarrow \BZ^d$ be an injective endomorphism, and let the sequence $(f_n)_{n \in \BZ^d}$ be ultimately periodic with respect to $H$. Let $F: \BZ^d \rightarrow \BZ^d$ be a different injective endomorphism satisfying the hypotheses of Theorem \ref{mainresult1} (including that $F$ is diagonalizable). Let $\Delta$ be the set of values that the sequence takes. Since $\BZ^d$ is a finitely generated abelian group and $H$ is an injective endomorphism on $\BZ^d$, $\BZ^d/H(\BZ^d)$ is finite. 

Suppose that $(f_n)_{n \in \BZ^d}$ is periodic with respect to $H$.  By Lagrange's Theorem, if $k = [\BZ^d: H(\BZ^d)]$, then $k\BZ^d \subseteq H(\BZ^d)$. Then $kI$ is an injective endomorphism as well, and so $(f_n)_{n \in \BZ^d}$ is periodic with respect to $kI$ as well. Let $Q$ be a set of representatives for the cosets of $\BZ^d/kI(\BZ^d)$. Choose a digit set $D$ according to Lemma \ref{digitsetlemma} so that every $x \in \BZ^d$ has a finite $(D, F)$ expansion. For $d \in D$ and $q \in Q$, define $\delta(q, d) = Fq + d \mod kI(\BZ^d)$, that is, transition to the coset of $\BZ^d/k\BZ^d$ containing $Fq + d$. This is well defined since $F(kI(\BZ^d)) \subseteq kI(\BZ^d)$, while the same cannot necessarily be said for $H$. Have $q_0$ be the representative $0$, and let the output function $\tau: Q \rightarrow \Delta$ be given by $\tau(q) = f_q$. We claim that $M = (Q, D, \delta, q_0, \Delta, \tau)$ is a DFAO for $(f_n)_{n \in \BZ^d}$ in base $(D, F)$. To see this, suppose that we have some input $w = \sum\limits_{j=0}^m F^jd_j$ with $d_j \in D$. After reading $d_m$, we transition from $q_0 = 0$ to $F(0) + d_m = d_m \mod kI(\BZ^d)$, and after reading the entire input, we will be in a state $q$ which is congruent to $\sum\limits_{j=0}^m F^jd_j \mod kI(\BZ^d)$, that is, that $x := w - q \in kI(\BZ^d)$. The machine outputs $f_q$, but $f_w = f_{q + x} = f_q$ by periodicity since $x \in kI(\BZ^d)$ so we have the correct output. \\

Now suppose that $(f_n)_{n \in \BZ^d}$ is ultimately periodic with respect to $H$ (and hence $kI$). Let $(g_n)_{n \in \BZ^d}$ be a periodic sequence which agrees with $(f_n)_{n \in \BZ^d}$ except at finitely many terms. Call these terms $v_1, \ldots, v_k$. For each $v_i$, choose a base $(D, F)$-expansion $[w_1 \cdots w_n]_F = v_i$. We may produce a DFA $M_i$ which accepts the set $\{x \in D^*: [x]_F = v_i\}$ via the same method as in the proof of Lemma \ref{expanddigitlemma}, namely deciding whether $[x]_F = [w]_F$, only this time the word $w$ is fixed instead of being an arbitrary element of $D^*$. From the periodic case, there is a DFAO $N$ which computes $(g_n)_{n \in \BZ^d}$. Consider the product automaton $\left(\prod\limits_{i=1}^k M_i\right) \times N$, with an output function given by 
\begin{align*}
\tau_{M \times N}(s_1, \ldots, s_k, t) = \begin{cases}
f_{v_i} \text{ if } \tau_{M_i}(s_i) = 1 \\
\tau_N(t) \text{ otherwise},
\end{cases}
\end{align*}
which is well-defined since the sets of inputs accepted by the $M_i$ is disjoint by definition. This DFAO computes $(f_n)_{n \in \BZ^d}$ since it agrees with $(g_n)_{n \in \BZ^d}$ except on $\{v_1, \ldots, v_k\}$, where we have manually modified the output.

If instead we are in dimension $d = 1$, the construction is easier. Suppose that $f$ is periodic on $(-\infty, a]$ and $[b, \infty)$, possibly with different periods. Every $F$ is multiplication by an integer $k$ with $|k| > 1$. Letting $D$ be the digit set, after reading all but the last $r$ digits of an input, we have something of the form $k^r x + z$, with $z$ having a $k$-expansion consisting of at most $r$ digits. We have that
\begin{align*}
|z| = |\sum\limits_{i=0}^{r-1} k^i d_i| \leq \sum\limits_{i=0}^{r-1} |k|^i|d_i| \leq \left(\frac{|k|^r-1}{|k|-1}\right)\max_{d \in D}|d| < |k|^r\max_{d \in D} |d|,
\end{align*}
and so if $|x| > \max\{|a|, |b|\} + \max_{d \in D}|d|$, then $|kx + d| \geq 2|x| - \max_{d \in D}|d| > \max\{|a|, |b|\} + \max_{d \in D}|d|$, and so we will end up in either $(-\infty, a]$ or $[b, \infty)$. Thus we keep track of whether our current value exceeds these bounds, and if it ever does, transition to a set of states where we keep track of the value modulo the LCM of the two periods. Otherwise, we check against the exact values of $f$, requiring finitely many states to do so in a bounded region. Finally, if $k$ is negative, we also keep track of whether we have read an odd or even number of digits to determine the sign at every intermediate step.
\end{proof}

Toward a proof of the forward direction of Theorem \ref{mainresult1}, we consider prefixes which end up in the same state in both the $F$ and $G$-automata. We require some notation for this. \\

\begin{defn}
Let $M = (Q, D, \delta, q_0, \Delta, \tau)$ be a DFAO taking inputs in base $(D, F)$, and let $q \in Q$. We define $L_{q, F} := \{w \in D^*: \delta(q_0, w) = q\}$ to be the set of all strings which when read as input by $M$ end up in state $q$. If $X \subseteq D^*$, we write $[X]_F$ to mean the elements of $\BZ^d$ which have a base $(D, F)$-expansion belonging to $X$. Then $[L_{q, F}]_F$ denotes the elements of $\BZ^d$ which have some base-$(D, F)$ expansion that ends in state $q$ after being passed as input to $M$. \\
\end{defn}

\begin{lemma}\label{pumpspan}
Let $M$ and $N$ be minimal DFAOs (in the sense of Myhill-Nerode) which compute a sequence $(a_n)_{n \in \BZ^d}$ in bases $(D_F, F)$ and $(D_G, G)$ respectively, where $F$ and $G$ satisfy the first two hypotheses of Theorem \ref{mainresult1}, and $D_F$ and $D_G$ satisfy the constraints of Lemma \ref{digitsetlemma}, so that every element of $\BZ^d$ has a finite $(D_F, F)$-expansion and a finite $(D_G, G)$-expansion. Suppose that a state $s$ in $M$ is chosen so that $[L_{s, F}]_F$ is infinite. Then there exists a state $t$ in $N$ (depending on $s$), $l \geq 1$, and finitely many pairs $x_1, y_1, \ldots, x_r, y_r \in [L_{s, F}]_F \cap [L_{t, G}]_G$ such that $\mathrm{span}_\BQ\{x_i - y_i\}$ is a non-trivial $F^l$-invariant subspace of $\BQ^d$. 
\end{lemma}

\begin{proof}
If $M$ is a DFAO which computes $(a_n)_{n \in \BZ^d}$ in base $(D_F, F)$, then there is a DFA $M'$ which accepts precisely inputs that end in state $s$ in $M$. By the pumping lemma for regular languages, there exist strings $a, b, c \in D_F^*$ such that $M'$ accepts $ab^nc$ for all $n \geq 0$. Since $[L_{s, F}]_F$ is infinite, we may choose $a, b$, and $c$ so that $\{[ab^nc]_F: n \geq 0\}$ is an infinite set which ends up in state $s$. \\

Consider that if $[ab^{n+1}c] = [ab^nc]$, then
\begin{align*}
F^{|b|(n+1) + |c|}([a]_F) + F^{|c|}([b^{n+1}]_F) + [c]_F = F^{n|b| + |c|}([a]_F) + F^{|c|}([b^n]_F) + [c]_F,
\end{align*}
and so
\begin{align*}
0 &= F^{|b|(n+1)+|c|}([a]_F) + F^{|b|n+|c|}([b]_F) - F^{|b|n+|c|}([a]_F) \\
&= F^{|b|}([a]_F) + [b]_F - [a]_F
\end{align*}
by injectivity of $F$. Equivalently, we have
\begin{align*}
(I - F^{|b|})([a]_F) = ([b]_F).
\end{align*}
Since $F$ has all eigenvalues greater than $1$ in modulus, $I - F^{|b|}$ is invertible, and so we obtain
\begin{align*}
[a]_F = (I - F^{|b|})^{-1}([b]_F),
\end{align*}
and so for a fixed $b$, this is only possible for $a$ with a specific value $[a]_F$. Trim $M'$ to form an automaton $M''$ consisting of states $q$ for which there is a path from $q_0$ to $q$ and a path from $q$ to $s$. Suppose that for every state $p$ in $M''$ which is part of a cycle, any word $a$ leading from $q_0$ to $p$ has the same value $[a]_F$. Then for any word $w$ passing through such a $p$, $[w]_F$ is determined by the tail of the word after leaving all cycle states, since the head will just have value $[a]_F$. But then this tail can visit at most finitely many states, and so there are finitely many such tails, which implies that $[L_{s, F}]_F$ is finite, a contradiction. Thus there exists two values of $a$ for which $\delta(q_0, a) = p$ but with $[a]_F$ distinct. For one of these, we must have $[ab^{n+1}c]_F \neq [ab^nc]_F$. 

Additionally, we have that
\begin{align*}
[b^j]_F &= \sum\limits_{i=0}^{j-1} F^{i|b|}([b]_F) \\
&= \left(\sum\limits_{i=0}^{j-1} F^{i|b|}\right)([b]_F) \\
&= \left(\sum\limits_{i=0}^{j-1} F^{i|b|}\right)\left((I - F^{|b|})([a]_F)\right) \\
&= (I - F^{j|b|})([a]_F) \text{ by telescoping,}
\end{align*}
and so $(I - F^{j|b|})^{-1}([b^j]_F) = [a]_F$ as well. Note that $I - F^{j|b|}$ is still invertible, as $F^{j|b|}$ does not have $1$ as an eigenvalue since $F$ does not. Thus $b$ and $b^j$ both have the same bad value $[a]_F$ for the cycle, and so we may replace $b$ by any power of $b$ without loss of generality.

Choose $b$ such that $\lambda^{|b|} > 2$. We claim that for any $m > n \in \BN$, $[ab^mc]_F \neq [ab^nc]_F$. Write $u = F^{|b|}([a]_F) + [b]_F - [a]_F$. By reverse triangle inequality, we have that
\begin{align*}
\|[ab^{n+2}c]_F - [ab^nc]_F\|_V &= \|([ab^{n+2}c]_F - [ab^{n+1}c]_F) + ([ab^{n+1}c]_F - [ab^nc]_F)\|_V \\
&\geq \left|\|[ab^{n+2}c]_F - [ab^{n+1}c]_F\|_V - \|[ab^{n+1}c]_F - [ab^nc]_F\|_V\right| \\
& = \left|\|F^{|b|(n+1) + |c|}(u)\|_V - \|F^{|b|n+|c|}(u)\|_V\right| \\
&= \left|\|F^{|b|}(F^{|b|n + |c|}(u)\|_V - \|F^{|b|n+|c|}(u)\|_V\right| \\
&= \left|\lambda^{|b|}\|(F^{|b|n + |c|}(u)\|_V - \|F^{|b|n+|c|}(u)\|_V\right| \\
&> \|F^{|b|n+|c|}(u)\|_V \\
&= \|[ab^{n+1}c]_F - [ab^nc]_F\|_V. \\
\end{align*}

Now suppose that $\|[ab^{n+m}c]_F - [ab^{n+1}c]_F\|_V \geq \|[ab^{n+2}c]_F - [ab^{n+1}c]_F\|_V$. We have that
\begin{align*}
\|[ab^{n+m}c]_F - [ab^nc]_F\|_V &= \|([ab^{n+m}c]_F - [ab^{n+1}c]_F) + ([ab^{n+1}c]_F - [ab^nc]_F)\|_V \\
&\geq \left|\|[ab^{n+m}c]_F - [ab^{n+1}c]_F\|_V - \|[ab^{n+1}c]_F - [ab^nc]_F\|_V\right| \\
&> \left|\|[ab^{n+2}c]_F - [ab^{n+1}c]_F\|_V - \|[ab^{n+1}c]_F - [ab^nc]_F\|_V\right| \\
&> \|[ab^{n+1}c]_F - [ab^nc]_F\|_V.
\end{align*}

By induction, $\|[ab^{n+m}c]_F - [ab^nc]_F\|_V > \|[ab^{n+1}c]_F - [ab^nc]_F\|_V$ for all $m, n \in \BN$, and so in particular all elements of $\{[ab^nc]_F: n \in \BN\}$ are pairwise distinct. 

We next claim that we may take $a, b, c$ to all be of some uniform length $\ell$. To see this, take $j$ large enough so that $|b^j| \geq |a| + |c|$. Then $\{[a(b^j)^nc]_F: n \geq 0\}$ is still infinite since all $[ab^nc]_F$ are distinct. Let this $b^j$ be our new $b$, and then prepend $a$ with enough zeroes so that $|b| = |a| + |c|$. Note that we may prepend by zeroes since $M$ was assumed to be minimal, and we have $[0w]_F = [w]_F$, so that these two words belong to the same Myhill-Nerode equivalence class, and so $\delta(q_0, 0) = q_0$. Write $b = xy$ with $|x| = |c|$ and $|y| = |a|$. Then $ab^{m+1}c = a(xy)^{m+1}c = (ax)(yx)^m(yc)$ with $|a| + |c| = |ax| = |yx| = |yc|$. Take $ax, yx$, and $yc$ to be our new $a, b$, and $c$ respectively, all of a uniform length. \\

Since every $[ab^nc]_F$ has a finite $(D_G, G)$-expansion, it must end in some state in $N$. Since there are finitely many states, the sets
\begin{align*}
\{[ab^nc]_F: n \geq 0\} \cap [L_{t, G}]_G
\end{align*}
form a finite cover of the infinite set $\{[ab^nc]_F: n \geq 0\}$ across all states $t$ of $N$. In particular, for some $t$, the set $\{[ab^nc]_F: n \geq 0\} \cap [L_{t, G}]_G$ must have positive upper density in $\{[ab^nc]_F: n \geq 0\}$. Then the set $\{n \in \BN: [ab^nc]_F \in [L_{t, G}]_G\}$ has positive upper density in $\BN$ as well. By Szemer\'edi's Theorem, this set contains arithmetic progressions of arbitrary length. Then for any $C \in \BN$, there exist $k, m \in \BN$ such that for $n < C$,
\begin{align*}
[ab^{kn + m}c]_F \in [L_{t, G}]_G.
\end{align*}

Given that $|a| = |b| = |c| = \ell$, for any $n \geq 0$ we have that
\begin{align*}
[ab^{k(n+1) + m}c]_F - [ab^{kn + m}c]_F &= F^{(k(n+1)+m+1)\ell}([a]_F) + F^{(kn+m+1)\ell}([b^k]_F) \\ 
& \quad{}- F^{(kn+m+1)\ell}([a]_F) \\
& = F^{(kn+m+1)\ell}(F^{k\ell}([a]_F) + [b^k]_F - [a]_F), 
\end{align*}
so that
\begin{align*}
F^{kl}([ab^{k(n+1)+m}c]_F - [ab^{kn+m}c]_F) = [ab^{k(n+2)+m}c]_F - [ab^{k(n+1)+m}c]_F.
\end{align*}
Let $P(x)$ be the characteristic polynomial of $F^{k\ell}$ over $\BQ$, of degree at most $d$, so that $P(F^{kl}) = 0$ by Cayley-Hamilton. Then $F^{(m+1)\ell}P(F^{kl}) = 0$, and so in particular, $F^{(m+1)\ell}P(F^{kl})(F^{kl}([a]_F) + [b^k]_F - [a]_F) = 0$. This means that some $\BQ$-linear combination of 
\begin{align*}
S := \{[ab^{k(n+1)+m}c]_F - [ab^{kn+m}c]_F: 0 \leq n \leq \deg P\}
\end{align*}
equals zero. In particular, taking $C > d+1$, we then have that
\begin{align*}
[ab^{k(n+1) + m}c]_F - [ab^{kn + m}c]_F \in \mathrm{span}_\BQ S
\end{align*}
for $\deg P < n < C$ as well. In particular, this means that $\mathrm{span}_\BQ S$ is a non-trivial $F^{k\ell}$-invariant subspace of $\BQ^d$. Set $l = k\ell$, and we are done.
\end{proof}

\begin{cor}\label{pumpspancor}
If in addition to the above premises, $F$ and $G$ also satisfy the third hypothesis of Theorem \ref{mainresult1}, then the set
\begin{align*}
W = \mathrm{span}_\BQ\{x - y: x, y \in [L_{s, F}]_F \cap [L_{t, G}]_G\}
\end{align*}
from the above lemma is equal to $\BQ^d$ itself.
\end{cor}

\begin{proof}
Without loss of generality, suppose there are no non-trivial $F^l$-invariant subspaces of $\BQ^d$ for any $l \geq 1$. If this is true for $G$ instead simply swap $F$ and $G$'s role in the argument of the above lemma. In this lemma, we remarked that $\mathrm{span}_\BQ S$ is a non-trivial $F^{k\ell}$-invariant subspace of $\BQ^d$, and so this must be $\BQ^d$ itself, and $W$ contains $\mathrm{span}_\BQ S$, so $W = \BQ^d$ as well. \\
\end{proof}

We are now ready to prove the main result of this section.

\begin{proof}[Proof of Theorem \ref{mainresult1}, forward direction]

By symmetry, we may assume that $F$ has no non-trivial invariant subspaces over $\BQ^d$ among its powers rather than $G$. Let $M_F$ and $M_G$ be automata which compute $(f_n)_{n \in \BZ^d}$ in bases $(D_F, F)$ and $(D_G, G)$ respectively. Let $\lambda$ be the common modulus of eigenvalues of $F$, and $\mu$ the common modulus of eigenvalues of $G$. Choose digit sets $D_F'$ and $D_G'$ in accordance with Lemma \ref{digitsetlemma}, and let $M_F' = (Q_F, D_F', \delta_F, q_{0, F}, \Delta, \tau_F)$ and $M_G' = (Q_G, D_G', \delta_G, q_{0, G}, \Delta, \tau_G)$ be the automata computing $(f_n)_{n \in \BZ^d}$ in bases $(D_F', F)$ and $(D_G', G)$ respectively given by Corollary \ref{expanddigitdfao}. Minimize these automata in the sense of Myhill-Nerode. Note that for some $q \in Q_F$, if we have $x, y \in [L_{q, F}]_F$ and $z \in [(D_F')^n]_F$ (meaning $z$ has a representation in base $(D_F', F)$ with at most $n$ digits), then $f_{F^n x + z} = f_{F^n y + z}$. The same is true on $[L_{q, G}]_G$ for any $q \in Q_G$ as well. \\

Let $Q_\infty$ be the set of $q \in Q_F$ such that $[L_{q, F}]_F$ is infinite. For each $q$, choose $r_q \in Q_G$ according to Lemma \ref{pumpspan} and Corollary \ref{pumpspancor} so that there is a basis $B = \{b_{q, 1}, \cdots, b_{q, d}\}$ for $\BQ^d$ such that each $b_{q, j} \in \{x-y: x, y \in [L_{q, F}]_F \cap [L_{r_q, G}]_G\}$. For $1 \leq j \leq d$, write $b_{q, j} = x_{q, j} - y_{q, j}$ with $x_{q,j}, y_{q,j} \in [L_{q,F}]_F \cap [L_{r_q,G}]_G$. We may write each standard basis vector $e_j$ as a $\BQ$-linear combination of elements from $B$, and through clearing all denominators, across all $j$, there is some constant $K_q \in \BZ$ such that each $K_q e_j$ belongs to the $\BZ$-module generated by $B$. Let $K = \prod\limits_{q \in Q_\infty} K_q$. Write
\begin{align*}
Ke_j = \sum\limits_{k=1}^d s_{q,j,k}b_{q,k},
\end{align*}
and then write
\begin{align*}
T_q = \sum\limits_{j=1}^d \sum\limits_{k=1}^d |s_{q,j,k}|(\|x_{q,k}\|_V + \|y_{q,k}\|_V + 1).
\end{align*}
Finally, let $T = \max\limits_{q \in Q_\infty} T_q$. \\

Now fix some $q \in Q_\infty$ and let $x \in [L_{q, F}]_F$. By Lemma \ref{closepowerslemma}, for any $\epsilon > 0$ we can find arbitrarily large $m, n \in \BN$ such that $\|F^m - G^n\|_V \leq \epsilon\min\{\lambda^m, \mu^n\}$. Let $L$ be the constant defined in the statement of Lemma \ref{digitsetlemma}. We have that
\begin{align*}
f_{F^m x + z} &= f_{F^m x_{q,1} + z} \text{ since } x, x_{q,1} \in [L_{q,F}]_F \\
&= f_{G^n x_{q,1} + z + (F^m - G^n)x_{q,1}} \\
&= f_{G^n y_{q,1} + z + (F^m - G^n)x_{q,1}} \text{ since } x_{q,1}, y_{q,1} \in [L_{r_q,G}]_G \\
&= f_{F^m y_{q,1} + z + (F^m - G^n)b_{q,1}} \\
&= f_{F^m x + z + (F^m - G^n)b_{q,1}},
\end{align*}
where these equalities depend on the remainders outside the images of $F^m$ and $G^n$ having expansions of length at most $m$ and $n$ respectively, since this ensures that these values get read into the automata as written. Because of how we defined $T$, we can apply this logic $s_{q,j,1}$ times to get that $f_{F^m x + z} = f_{F^m x + z + s_{q,j,1}(F^m - G^n)b_{q,1}}$. If $s_{q,j,1}$ is negative, simply swap instances of $x_{q,1}$ and $y_{q,1}$ in the argument above. We can also do this for all $b_{q,k}$ with $1 \leq k \leq d$, and so in particular, we may obtain that $f_{F^m x + z} = f_{F^m x + z + (F^m - G^n)Ke_j}$. We may further combine these to establish that $f_{F^m x + z} = f_{F^m x + z + (F^m - G^n)p}$ for any $p = Kv$ with $\|v\|_\infty = 1$. Take $z$ to be such that $\|z\|_V \leq (1-\delta)L\min\{\lambda^m, \mu^n\}$ for some $\delta$ small. In the worst case, we have that
\begin{align*}
\left\|z + (F^m - G^n)K\sum\limits_{j=1}^d e_j\right\|_V &\leq \|z\|_V + \|F^m - G^n\|_V\left\|\sum\limits_{j=1}^d\sum\limits_{k=1}^d s_{q,j,k}b_{q,k}\right\|_V \\
&\leq (1-\delta)L\min\{\lambda^m, \mu^n\} + \epsilon\min\{\lambda^m, \mu^n\}\left(\sum\limits_{j=1}^d\sum\limits_{k=1}^d |s_{q,j,k}|(\|x_{q,k}\|_V + \|y_{q,k}\|_V)\right) \\
&< (1-\delta)L\min\{\lambda^m, \mu^n\} + \epsilon T_q\min\{\lambda^m, \mu^n\} \\
&\leq (1-\delta)L\min\{\lambda^m, \mu^n\} + \epsilon T\min\{\lambda^m, \mu^n\} \\
&\leq L\min\{\lambda^m, \mu^n\} \text{ for sufficiently small } \epsilon,
\end{align*}
which still has a $(D_F', F)$-expansion of at most $m$ digits and a $(D_G', G)$-expansion of at most $n$ digits, and so all intermediary equivalences are valid. \\ 

Since $q$ was an arbitrary state in $Q_\infty$, this holds for all $x \in [L_{q, F}]_F$ for some $q \in Q_\infty$. There are finitely many $x \in \BZ^d$ which do not fit the above, so let $\alpha$ be large enough so that $\{x \in \BZ^d: \|x\|_\infty \geq \alpha\}$ avoids them. Define
\begin{align*}
B_V(x, R) = \{y \in \BC^d: \|x - y\|_V \leq R\},
\end{align*}
Let $c_0 \in \BZ$ be large enough so that $\{x \in \BC^d: \|x\|_V \leq L\} \subseteq \{x \in \BC^d: \|x\|_\infty \leq c_0\}$. Taking $m$ and $n$ as above, let $P_m' := \{w \in \BZ^d: \|F^{-m}w\|_\infty < \alpha + c_0\}$. 

We showed above that $f$ is periodic on $B_V(F^m x, (1-\delta)L\min\{\lambda^m, \mu^n\})$. As $F^m x$ for $x \in \BZ^d$ forms a lattice, the maximum distance possible from a point in this lattice corresponds to the center of the fundamental parallelipiped, which has distance
\begin{align*}
\frac{1}{2} \max_{\sigma \in \{\pm 1\}^d} \left\|F^m\left(x + \sum\limits_{j=1}^d \sigma_j e_j\right) - F^m(x)\right\|_V &= \frac{1}{2}\max_{\sigma \in \{\pm 1\}^d}\left\|F^m\left(\sum\limits_{j=1}^d \sigma_j e_j\right)\right\|_V \\
&\leq \frac{1}{2}\lambda^m \max_{\sigma \in \{\pm 1\}^d}\left\|\sum\limits_{j=1}^d \sigma_j e_j\right\|_V \\
&\leq \frac{1}{2}L\lambda^m,
\end{align*}
and $\frac{1}{2}L\lambda^m < (1-\delta)L\min\{\lambda^m, \mu^n\}$ for appropriately chosen $m$ and $n$ by Lemma \ref{closepowerslemma}.
Suppose $w \not \in P_m'$. Analogous to the above, if $x \in \BZ^d$ is the nearest integer vector to $F^{-m}w$ in $V$-norm, we have that $\|F^{-m}w - x\|_V \leq \frac{1}{2}L$, and so $\|F^{-m}w - x\|_\infty \leq c_0$, so that $\|x\|_\infty \geq \|F^{-m}w\|_\infty - c_0 \geq \alpha$, so $x$ is not a bad point. If we write $w = F^m x + z$, then $\|z\|_V = \|F^m(F^{-m}w - x)\|_V \leq \frac{1}{2}L\lambda^m$, and so $w$ lies close to the center of a periodicity ball. From this, we have periodicity outside of $P_m'$ in any direction $K(F^m - G^n)x$ with $x \in \BZ^d$ and $\|x\|_\infty = 1$. 

If $d \geq 2$, we claim that there is a unique way to assign values to $f$ on $P_m'$ to form a periodic sequence. Suppose not, so that there is a point $z \in P_m'$ and $x, y \in \BZ^d\setminus P_m'$ such that $z = x + K(F^m - G^n)u$ and $z = y + K(F^m - G^n)v$ for some $u, v \in \BZ^d$, but $f_x \neq f_y$. Then $y = x + K(F^m - G^n)(u-v)$. If the direct route from $x$ to $y$ passes through $P_m'$, we claim that we may avoid it by going around. Note that $K(F^m - G^n)$ is injective, since no eigenvalue of $F^m - G^n$ can vanish, as then we would have $\lambda^m e^{im\theta_j} = \mu^n e^{in\psi_j}$ and hence $\lambda^m = \mu^n$, contradicting the multiplicative independence of $\lambda$ and $\mu$. We want to find a sequence of $K(F^m - G^n)e_j$ such that traveling according to these vectors avoids $P_m'$. We have that $x + K(F^m - G^n)z \in P_m' \iff K(F^m - G^n)^{-1}(x) + z \in K(F^m - G^n)^{-1}(P_m')$, and $K(F^m - G^n)^{-1}(P_m')$ is still a bounded region. Since $d \geq 2$, this region we are traveling over is connected, and so we may choose a sequence of $\pm e_j$ movements which net travels in the direction $u-v$ and avoids this bounded region by traveling far out in each direction initially. Each intermediary vector $z$ in this sequence satisfies $f_{x + K(F^m - G^n)z} = f_x$ by periodicity, and so in the end we have $f_x = f_y$, a contradiction. Hence we can continue $f|_{\BZ^d\setminus P_m'}$ to a periodic sequence $(p_n)_{n 
 \in \BZ^d}$ which differs from $(f_n)_{n \in \BZ^d}$ on finitely many terms, and so $(f_n)_{n \in \BZ^d}$ is ultimately periodic.

If instead we have $d = 1$, the above argument fails, but we have still established periodicity on $(-\infty, -\lambda^m(\alpha + c_0)]$ and $[\lambda^m(\alpha + c_0), \infty)$, albeit possibly with different periods. 
\end{proof}

\begin{cor}
Let $F, G \in M_{d \times d}(\BZ)$ satisfy the hypotheses of Theorem \ref{mainresult1}. Suppose that a set $A \subseteq \BZ^d$ is both $(D_F, F)$ and $(D_G, G)$-automatic. Then if $d \geq 2$, there exists a periodic set $P \subseteq \BZ^d$ with period $H$ for some injective endomorphism $H: \BZ^d \rightarrow \BZ^d$ such that $A \Delta P$ is finite.
\end{cor}

\begin{proof}
Let $(\chi_{A, n})_{n \in \BZ^d}$ be the characteristic sequence of $A$. By Theorem \ref{mainresult1}, this sequence is ultimately periodic. Then there exists some periodic binary sequence $(p_n)_{n \in \BZ^d}$ with period $H$ for some injective endomorphism $H: \BZ^d \rightarrow \BZ^d$ which differs from $(\chi_{A, n})_{n \in \BZ^d}$ on finitely many terms. Let $P$ be the set whose characteristic sequence is $(p_n)_{n \in \BZ^d}$. Then $P$ is a periodic set with period $H$, and $A \Delta P$ is finite.
\end{proof}

We end this section with a counterexample showing that the conclusion of Theorem \ref{mainresult1} in the case of $d \geq 2$ becomes false in the case of $d = 1$, and that it is not merely excluded due to the proof technique used. \\

\begin{ex}
With $d = 1$, let $F$ be multiplication by $2$, $G$ multiplication by $3$, and $D_F = D_G = \{-1, 0, 1\}$. Then $\BN$ is $(F, D_F)$ and $(G, D_G)$-automatic since in such numeration systems, whether a number is negative is determined by whether its leading non-zero digit is negative. Also, $2\BZ$ is $(F, D_F)$-automatic by checking whether the last digit is $0$, and $(G, D_G)$-automatic by computing the sum of the digits modulo $2$. Then $A = \BN \cup 2\BZ$ is both $(F, D_F)$ and $(G, D_G)$-automatic as well, but is not ultimately periodic with respect to any non-zero period $p$, since $A$ contains all sufficiently large positive integers, and hence so does $P$. But $P$ is determined modulo $p$, so we have $P = \BZ$, and yet $A$ omits every negative odd number, so that $A \Delta P$ is not finite.
\end{ex}

\section{Allowing for Invariant Subspaces}
We wish to eliminate the artificial requirement that either there are no non-trivial $F^\ell$-invariant subspaces of $\BQ^d$ or no non-trivial $G^\ell$ invariant subspaces of $\BQ^d$ for any $\ell > 0$. Not only can one not expect this to be true for a general matrix $F$, but it is also difficult to prove in practice. We present the following alternative instead: \\

\begin{thm}\label{mainresult2}
Let $F, G \in M_{d \times d}(\BZ)$ and let $D_F, D_G \subseteq \BZ^d$ be digit sets. Suppose that the following assumptions hold:

\begin{itemize}
\item $F$ and $G$ commute and are diagonalizable over $\BC$ (equivalently, they are simultaneously diagonalizable).
\item All of the eigenvalues of $F$ over $\BC$ have the same modulus $\lambda > 1$, and all of the eigenvalues of $G$ over $\BC$ have the same modulus $\mu > 1$, with $\lambda$ and $\mu$ being multiplicatively independent.
\item $F$ has $d$ distinct eigenvalues, and $\lambda_i/\lambda_j$ is not a root of unity whenever $i \neq j$.
\end{itemize}

Then if a set $A$ is both $(D_F, F)$ and $(D_G, G)$-automatic, it is Presburger definable. \\
\end{thm}

To prove this, we will need a different approach. In \cite{Muchnik}, Muchnik uses the logical formulation of the Cobham-Semenov Theorem, and the following criterion for Presburger definability: \\

\begin{thm}[Theorem 4 of \cite{Muchnik}]
If all sections of a set $A \subseteq \BZ^d$ are definable and there exists a finite set $V$ of non-zero vectors such that $A$ is periodic in every far enough neighbourhood of radius $k$ (where $k$ is the sum of the lengths of vectors in $V$) with a period from $V$, then $A$ is definable in Presburger arithmetic. \\
\end{thm}

Here the $(i, \ell)$-section of $A$ is defined to be 
\begin{align*}
\{(x_1, \ldots, x_{i-1}, x_{i+1}, \ldots, x_d): (x_1, \ldots, x_{i-1}, \ell, x_{i+1}, \ldots, x_d) \in A\},
\end{align*}
and $V$ is called the \textit{set of possible periods of $A$}. \\

Unfortunately, this proof does not easily extend to the setting of $F$-automatic sets. For one, $F$ need not preserve sectioning out by coordinates, and so the sections of $A$ need not be $F$-automatic themselves. Instead, we will want to define what it means to section out by an arbitrary subspace. Even then, having a finite set of possible periods will not be enough, since a single vector often will not span the subspace by which we have sectioned out. Finally, unlike in the simple case of sectioning out by a coordinate, this does not easily reduce things to the same setting but with $d-1$ coordinates instead of $d$. \\

Our approach will be as follows: we will section out by $F$-invariant subspaces $W$, which were the things forbidden in the proof of Lemma \ref{pumpspan}, and which now possibly occur due to having discarded the third assumption in the statement of Theorem \ref{mainresult1}. We show that a generalization of Muchnik's characterization of Presburger definability holds in this setting, where instead of having finitely many possible period vectors, we will instead have finitely many possible \textit{period bases} each of which generates some $F$-invariant subspace $W$, and for which we are periodic on all basis elements. We use this criterion to reduce the problem to proving that the sections of $A$ with respect to $W$ are Presburger definable, and we show that since $W$ is $F$-invariant, these sections have $F$-automata and $G$-automata and lie in subspaces of strictly lesser dimension. By descent, we will eventually reach dimension $d=1$, after which we may appeal to Theorem \ref{mainresult1} to conclude the proof. One might note that to use this theorem, we need there to be no $F^\ell$-invariant subspaces for any $\ell \geq 1$. We will show that with the third assumption in the statement of our theorem, $F$-invariance and $F^\ell$-invariance are equivalent. We note that when we speak of quantifying over ``all $F$-invariant subspaces $W$ of $\BQ^d$", we are in fact quantifying over a finite set: since $F$ has $d$ distinct eigenvalues, there are at most $2^d$ such invariant subspaces.

We begin by outlining the correct definition of a generalized Muchnik's criterion.

\begin{defn}
Let $\mathcal{B}$ be a finite set of bases for various non-trivial $F$-invariant subspaces of $\BQ^d$. We say that a set $A$ is periodic with respect to a vector $u$ in a radius of $k$ around $x$ if $y \in A \iff y + u \in A$ whenever $y$ and $y+u$ both lie in the ball of radius $k$ around $x$ (with respect to $\|.\|_\infty$).

We say that a set $A$ is $\mathcal{B}$-periodic if for every $k$, there exists $\ell$ such that for each $x \in \BZ^d$ with $\|x\|_\infty > \ell$, there is a basis $\{u_1, \ldots, u_r\} \in \mathcal{B}$ such that $A$ is periodic with respect to each $u_i$ in a radius of $k$ around $x$. 
\end{defn}

Since we will end up reducing the problem to the sections of $A$, we would like the sections of $A$ to have $G$-automata which recognize them as well, and so we would like $W$ to be $G$-invariant. With the addition of the third assumption in the statement of Theorem \ref{mainresult2}, this is always possible. \\

\begin{lemma}\label{FequalsGinvariant}
Let $F$ satisfy the assumptions of Theorem \ref{mainresult2} which do not specify a $G$. Then if a subspace $W$ is $F^\ell$-invariant for some $\ell \geq 1$, it is $F$-invariant. Further, if $F$ and $G$ satisfy the assumptions of Theorem \ref{mainresult2} and if $W$ is $F$-invariant, it is $G$-invariant.
\end{lemma}

\begin{proof}
Since for any pair of eigenvalues $\lambda_i, \lambda_j$ of $F$ we have that $\lambda_i/\lambda_j$ is never a root of unity, for every $\ell$ we have that $\lambda_i^\ell \neq \lambda_j^\ell$, so $F^\ell$ has distinct eigenvalues as well (hence is diagonalizable), and the eigenspaces, which are all linear, coincide. First, we suppose that $W$ is a subspace over $\BC^d$. If $W$ is an invariant subspace of $F^\ell$, then $F^\ell|_W$ is also diagonalizable, and the eigenspaces of $F^\ell|_W$ are those of $F^\ell$ intersected with $W$, hence consisting of eigenlines as well. Each of these is clearly $F$-invariant, and so $W$ is as well. Now suppose $W$ is $F$-invariant. Then since $F$ and $G$ commute, they preserve each others eigenspaces. But the eigenvalues of $F$ are pairwise distinct, so these are again lines, and so $W$, being made up of such lines, is $G$-invariant as well. Now suppose instead that $W$ is a subspace over $\BQ^d$. Since extension of scalars commutes with linear maps, invariance of $W$ over $F$ and $G$ follows from invariance of $W \otimes \BC$. 
\end{proof}

Since $F$ has no repeated eigenvalues, its minimal polynomial is squarefree. Then $\BQ^d$ is a semisimple $\BQ[F]$-module, and so for any $F$-invariant rational subspace $W$ there exists a rational complement $W^c$ of $W$ which is also $F$-invariant. However, it does not follow that $\BZ^d = (\BZ^d \cap W) \oplus (\BZ^d \cap W^c)$. Instead this has finite index in $\BZ^d$, and we will need to keep track of coset representatives as well. We note that while it is possible to find a $\BZ$-module complement of $W$, such a complement would not necessarily be $F$-invariant, which is needed to prove Lemma \ref{secinherit}. We write $W_\BZ$ to mean $W \cap \BZ^d$. \\

\begin{defn}\label{sectiondef}
Let $A \subseteq \BZ^d$, and let $W$ be an $F$-invariant subspace of $\BQ^d$. Fix an $F$-invariant complement $W^c$ of $W$ so that $\BQ^d = W \oplus W^c$. Then viewed as $\BZ$-modules, we have that $[\BZ^d: W_\BZ \oplus (W^c)_\BZ]$ is finite. Fix coset representatives $t_1, \ldots, t_N$, and let $\ell \in W_\BZ$. We call the set $\{w \in (W^c)_\BZ: \ell + t_k + w \in A\}$ the $(\ell, t_k)$th $W$-subspace section of $A$, denoted $\Sec_{W, \ell, t_k}(A)$. Note that any $w \in W^c$ with $\ell + t_k + w \in A$ is always in $(W^c)_\BZ$ since $\ell, t_k \in \BZ^d$ and $A \subseteq \BZ^d$, so we are not missing anything by specifying this. \\
\end{defn}

With the above lemma and definition, we also have the machinery that we need in order to show that subspace sections of invariant subspaces are also automatic. \\

\begin{lemma}\label{secinherit}
For any $F$-invariant subspace $W$ of $\BQ^d$, any $\ell \in W_\BZ$, and any coset representative $t_k$ of $\BZ^d/(W_\BZ \oplus (W^c)_\BZ)$, if $A \subseteq \BZ^d$ is $F$-automatic, then $\Sec_{W, \ell, t_k}(A)$ is $F|_{W^c}$-automatic.
\end{lemma}

\begin{proof}
In this proof, we adopt the convention that we read from right to left.  $F|_{W^c}: W^c \rightarrow W^c$ is still an endomorphism, still injective since $F$ is, and $(W^c)_\BZ$ is still a finitely generated abelian group. Then all the necessary criteria are satisfied for the existence of a digit set $D_{W^c}$ within $W^c$ which gives rise to a complete numeration system using $F|_{W^c}$ on $W^c$, via Lemma \ref{digitsetlemma}. Since $A$ is $F$-automatic, there exists an $F$-automaton $M$ with input from digit set $D$ and which recognizes $A$. In particular, since $\ell, t_k \in \BZ^d$, $\ell$ and $t_k$ have base $F$ expansions with digits from $D$. Let $D' \supseteq D$ be large enough to contain $D + D + D_{W^c}$ (as a sum of sets). Then there is an $F$-automaton $M'$ with digit set $D'$ which recognizes $A$. From this, we may define an $F|_{W^c}$-automaton with digit set $D_{W^c}$ which recognizes $\{w \in (W^c)_\BZ: \ell + t_k + w \in A\}$ for fixed vectors $\ell$ and $t_k$: given an input $\sum\limits_{i=0}^m F^id_i$ with $d_i \in D_{W^c}$ for all $i$, write $\ell$ as its expansion $\sum\limits_{i=0}^m F^ib_i$ (possibly with leading zeroes) and similarly write $t_k$ as $\sum\limits_{i=0}^m F^ic_i$. Then we may feed $\sum\limits_{i=0}^m F^i(b_i + c_i + d_i)$ into $M'$, since $b_i + c_i + d_i \in D'$, and this will accept if and only if $\sum\limits_{i=0}^m F^id_i$ is in $\Sec_{W, \ell, t_k}(A)$. Note that we are implicitly using states here to keep track of when we have exceeded the lengths of one of these digit expansions without leading zeroes (and padding with them whenever appropriate), but this is fine since this minimum is bounded above by the lengths of $\ell$ and $t_k$, which are fixed in advance. 
\end{proof}

We define the boundary of $A$ in the direction $u \in \BZ^d$ to be \begin{align*}
\Bd(A, u) := \{x \in A: x + u \not \in A\}.
\end{align*}
We will prove several lemmas concerning these boundaries that will assist us in proving a generalized Muchnik's criterion.

\begin{lemma}\label{boundarysection}
If for every $F$-invariant subspace $W$, every $\ell \in W_\BZ$, and every coset representative $t_k \in \BZ^d/(W_\BZ \oplus (W^c)_\BZ)$ we have that $\Sec_{W, \ell, t_k}(A)$ is definable, then so is every $\Sec_{W, \ell, t_k}(\Bd(A, u))$ for each $u \in \BZ^d$.
\end{lemma}

\begin{proof}
Let $W$ be any $F$-invariant subspace. Since $\BQ^d = W \oplus W^c$, we have that $[\BZ^d: W_\BZ \oplus (W^c)_\BZ] = N' < \infty$, so let $s_1, \ldots, s_{N'}$ be coset representatives. Consider $\Sec_{W, \ell, s_k}(\Bd(A, u))$ for some $\ell \in W_\BZ$ and coset representative $s_k$. This is $w \in (W^c)_\BZ$ such that $w + \ell + s_k \in \Bd(A, u)$, which occurs when $w + \ell + s_k \in A$ but $w + \ell + s_k + u \not \in A$. We may write $u = u' + u'' + s_p$ for some $p$, with $u' \in W_\BZ$ and $u'' \in (W^c)_\BZ$. Writing $s_k + s_p = s'_W + s'_{W^c} + s_i$ for some coset representative $s_i$, and with $s'_W \in W_\BZ$ and $s'_{W^c} \in (W^c)_\BZ$, this can be characterized as
\begin{align*}
w \in \Sec_{W, \ell, s_k}(A) \land w + u'' + s'_{W^c} \not\in \Sec_{W, \ell + u' + s'_W, s_i}(A),
\end{align*}
a conjunction of two predicates that are definable by hypothesis. 
\end{proof}

\begin{lemma}\label{boundaryreduction}
Let $\mathcal{B}$ be a finite set of integral bases for various non-trivial $F$-invariant subspaces over $\BQ^d$ with $|\mathcal{B}| \geq 2$. Suppose that $A$ is $\mathcal{B}$-periodic and suppose that $\{u_1, \ldots, u_r\}$ is a basis in $\mathcal{B}$. Then for any $u \in \{\pm u_1, \ldots, \pm u_r\}$, $\Bd(A, u)$ is $\mathcal{B}\setminus\{\{u_1, \ldots, u_r\}\}$-periodic.
\end{lemma}

\begin{proof}
Let $T_\mathcal{B}$ be the highest infinity norm of any vector from any basis in $\mathcal{B}$. Let $k$ be arbitrary, and pick $\ell$ so that for every $\|x\|_\infty > \ell$, $A$ has some period basis $\{w_1, \ldots, w_s\}$ from $\mathcal{B}$ in a $k + T_\mathcal{B}$ radius neighbourhood of $x$. If $u = \pm w_i$ for some $i$, then if $y$ lies in a radius $k$ neighbourhood of $x$, $y + u$ lies in a neighbourhood of radius $k + T_\mathcal{B}$, and so $y \in A \iff y + u \in A$. By periodicity, $\Bd(A, u)$ is empty in a $k$-radius neighbourhood of $x$, so that it is vacuously periodic with respect to any basis from $\mathcal{B}\setminus\{\{u_1, \ldots, u_r\}\}$. If instead $u \neq \pm w_i$ for all $i$, suppose that $y$ and $y+w_i$ both lie in a neighbourhood of $x$ of radius $k$. Then if $y \in \Bd(A, u)$, we have $y \in A$ and $y + u \not \in A$. Since $y \in A$ and $y, y+w_i$ both lie in a neighbourhood around $x$ of radius $k < k + T_\mathcal{B}$, $y+w_i \in A$ as well. Then $(y + w_i) + u = (y + u) + w_i \not \in A$ since $y+u \not \in A$ and both lie in a neighbourhood of radius $k+T_\mathcal{B}$ around $x$, and thus $y+w_i \in \Bd(A, u)$ as well. The same argument applies in the other direction as well, using $-w_i$ instead of $w_i$. So $\Bd(A, u)$ is periodic with respect to this basis in a $k$-radius neighbourhood of $x$, and since $u \in \{\pm u_1, \ldots, \pm u_r\}$, we have that $\{w_1, \cdots, w_s\} \neq \{u_1, \ldots, u_r\}$.
\end{proof}

\begin{lemma}\label{boundarystep}
Let $U$ be an $F$-invariant subspace of $\BQ^d$, and let $\{u_1, \ldots, u_r\}$ be an integral basis for $U$. Then $A$ is definable in terms of $\Bd(A, u_i)$ and $\Bd(A, -u_i)$ for $1 \leq i \leq r$ as well as a finite number of sections of $A$ with respect to the subspace $U$.
\end{lemma}

\begin{proof}
Suppose $x \in A$, and let
\begin{align*}
S_x = \{(q_1, \ldots, q_r) \in \BZ^r: x + \sum\limits_{i=1}^r q_iu_i \in A\}.
\end{align*}
Clearly $0 \in S_x$. If $x \in \Bd(A, \pm u_j)$ for some $j$, this is clearly definable, so we assume this is not the case.

First suppose that $S_x \neq \BZ^r$. Then there is some bad tuple $(q_1, \ldots, q_r)$ such that $x + \sum\limits_{i=1}^r q_iu_i \not \in A$. We explain the case where $(q_1, \ldots, q_r) \in \BN^r$: there will be $2^r - 1$ other cases with similar formulas by replacing $u_i$ with $-u_i$ and vice versa for indices $i$ on which we have negative $q_i$. We may walk towards this tuple, augmenting first by $u_1$ until we have done so $q_1$ times, then by $u_2$ until we have done so $q_2$ times, and so forth up to $u_r$ augmented $q_r$ times. At some point, for the first time, we will reach some first point in this path for which we are no longer in $A$. We claim that this is definable in terms of $\Bd(A, \pm u_i)$ across all $i$. We wish to say that there exists $y \in \Bd(A, u_{r'})$ such that $y - x = \sum\limits_{i=1}^{r'} q_i'u_i$ for some $(q_1', \ldots, q_r') \in \BN^r$ and that every intermediary vector along the walk from $x$ to $y$ (including the two endpoints) does not lie in $\Bd(A, -u_i)$ for the respective relevant index $i$. We let $y$ be the last point on the walk lying in $A$ before we leave $A$ for the first time, so that in particular we have $y \in \Bd(A, u_{r'})$ where $u_{r'}$ is the direction of the first step out of bounds. To see that this implies that $x \in A$, note that $y = x + \sum\limits_{i=1}^{r'} q_i'u_i \in \Bd(A, u_{r'})$ means in particular that $x + \sum\limits_{i=1}^{r'} q_i'u_i \in A$, while $x + \sum\limits_{i=1}^{r'} q_i'u_i \not \in \Bd(A, -u_{r'})$ means that either $x + \sum\limits_{i=1}^{r'} q_i'u_i \not \in A$ or $x - u_{r'} + \sum\limits_{i=1}^{r'} q_i'u_i \in A$, and so it must be the latter. Repeating these implications, we descend and get that each step along the walk from $x$ to $y$ lies in $A$, including $x$ itself. 

Now suppose instead that $S_x = \BZ^r$. Here we cannot simply assert that $x + \sum\limits_{i=1}^r q_iu_i \not \in \Bd(A, \pm u_j)$ for all $(q_1, \ldots, q_r) \in \BZ^r$ and all $j$, since this also holds for the scenario where all of these points are not in $A$ as well. Choose $(q_1, \ldots, q_r)$ such that the $U$-component of $x + \sum\limits_{i=1}^r q_iu_i$ lies in a fixed bounded box, which is possible since $\{u_1, \ldots, u_r\}$ is a basis of $U$, and thus the $U$-component can be reduced modulo $\mathrm{span}_\BZ\{u_1, \ldots, u_r\}$. So there are finitely many $\ell \in U_\BZ$ and finitely many coset representatives $t_k$ such that $y = x + \sum\limits_{i=1}^r q_iu_i \in A$ if and only if the $(U^c)_\BZ$-component lies in some $\Sec_{U, \ell, t_k}(A)$, and no intermediary point on the walk from $x$ to $y$ lies in $\Bd(A, -\mathrm{sgn}(q_i)u_i)$ for its respective index. As before, a series of implications gives us that $x \in A$ as well. 
\end{proof}

The above is sufficient to prove a result in the style of Muchnik's criterion, though we take sections with respect to an $F$-invariant subspace rather than a fixed coordinate. \\

\begin{prop}\label{muchnikcriterion}
Let $A \subset \BZ^d$. Suppose that $\Sec_{U, v, t_k}(A)$ is Presburger definable for every non-trivial $F$-invariant subspace $U$ of $\BQ^d$, every $v \in U_\BZ$, and every coset representative $t_k$ of $\BZ^d/(U_\BZ \oplus (U^c)_\BZ)$, and suppose that there exists a finite set of possible bases $\mathcal{B}$ such that $A$ is $\mathcal{B}$-periodic. Then $A$ is Presburger definable.
\end{prop}

\begin{proof}
We proceed by induction on the number of possible bases within $\mathcal{B}$. For the base case, suppose there is a single possible basis $\{u_1, \ldots, u_r\}$ for $U = \mathrm{span}_\BQ\{u_1, \ldots, u_r\}$, with $[\BZ^d : U_\BZ \oplus (U^c)_\BZ] = N$ as $\BZ$-modules, and with $\{u_{r+1}, \ldots, u_d\}$ being a corresponding basis for $U^c$. Let $t_1, \ldots, t_N$ be coset representatives. Given a vector $x$, we may write it in terms of $U$ and $U^c$-coordinates by applying some matrix $E$. Since the entries of $E$ are constants determined by our choice of submodule $U$ and our basis for $U$, they are known in advance, and so this is Presburger definable. Since $A$ is $\mathcal{B}$-periodic, setting $k = \max\limits_{1 \leq i \leq r} \|u_i\|_\infty$, we may find $\ell > 0$ such that that $\{x \in \BZ^d: \|x\|_\infty > \ell\}$ is periodic with respect to $u_i$ for $1 \leq i \leq r$, and hence periodic with respect to any $\BZ$-combination of these vectors reachable by a path which stays in this region. Additionally, there is an $m$ so that if in the norm with respect to our new coordinates $\|Ex\|_\infty > m$, then $\|x\|_\infty > \ell$ in terms of the old coordinates. The region defined by $\|Ex\|_\infty \leq m$ is bounded, hence finite, and thus is clearly definable. For $\|Ex\|_\infty > m$, there are $3^r$ cases to consider, determined by whether $(Ex)_i > m$, $(Ex)_i < -m$, or $-m \leq (Ex)_i \leq m$ for $i = 1, \ldots, r$. Here $(Ex)_i$ represents the value of the $i$th of our new coordinates after changing basis with $E$. For each such case, if $I$ is the set of indices for which $(Ex)_i > m$ and $J$ is the set for which $(Ex)_i < -m$, then by taking steps in each $u_i$ direction while still satisfying $\|Ex\|_\infty > m$ until we can't anymore, there exist $q_i \in \BN$ for $i \in I \cup J$ such that $m < (E(x - \sum_{i \in I} q_iu_i + \sum_{i \in J} q_ju_j))_i < m + \sum_{1 \leq i \leq r} \|u_i\|_\infty$ for $i \in I$ and the negative version holds for $i \in J$. Of course, for values of $i$ in neither $I$ nor $J$, the value of the coordinate is strictly between $-m$ and $m$. For each $i$, there are finitely many $v \in U_\BZ$ which satisfy $-m < v_j < m + \sum\limits_{1\leq i\leq r} \|u_i\|_\infty$ for each of $1 \leq j \leq r$, and each of these corresponds with some coset representative $t_k$. There are finitely many $U$-sections $\Sec_{U, v, t_k}(A)$ which correspond to these possibilities, and each of these are definable. Across all $3^r$ cases, this is still finite. \\

For the inductive step, take some finite $\mathcal{B}$ with $|\mathcal{B}| \geq 2$, and suppose the statement holds for all sets of possible bases with fewer members. Let $\{u_1, \ldots, u_r\}$ be a basis from $\mathcal{B}$, and let $U = \mathrm{span}_\BQ\{u_1, \ldots, u_r\}$. By Lemma \ref{boundarystep}, we have that $A$ is definable in terms of $\Bd(A, u_i)$ and $\Bd(A, -u_i)$, as well as finitely many sections of $A$ with respect to $U$, the latter of which are definable by assumption. Let $u = \pm u_i$ for some $i$. By Lemma \ref{boundaryreduction}, $\Bd(A, u)$ is $B\setminus\{\{u_1, \ldots, u_r\}\}$-periodic. This set has fewer possible bases, and by Lemma \ref{boundarysection}, all sections of $\Bd(A, u)$ are definable. Hence by the inductive hypothesis, $\Bd(A, u)$ is definable, and thus $A$ is definable as well.
\end{proof}

\begin{rem}\label{muchnikremark}
In the above, it suffices to consider periodicity in neighbourhoods of radius equal to $|\mathcal{B}|T_\mathcal{B}$. This is because in the base case of the induction, it sufficed to have periodicity in neighbourhoods of radius $k = \max\limits_{1 \leq i \leq r} \|u_i\|_\infty$, where $\{u_1, \ldots, u_r\}$ is our only basis, and in the inductive step, if $A$ is periodic in neighbourhoods with radius $k$, then $\Bd(A, u)$ is periodic in neighbourhoods with radius $k - T_B$.

\end{rem}

In order to prove the main result, the last thing we need is to be able to reduce the problem from $A$ to the sections $\Sec_{W, \ell, t_k}(A)$. We would like to work over $(W^c)_\BZ$ of lesser dimension using $F|_{W^c}$ and $G|_{W^c}$. To do this, we require a transfer lemma. \\

\begin{lemma}\label{transferlemma}
Let $W$ be an $F$-invariant subspace of $\BQ^d$ with $W^c$ its invariant complement. Let $A' \subseteq (W^c)_\BZ$, and put $d' := d - \dim W$. Then there exists an isomorphism $\Phi: \BZ^{d'} \rightarrow (W^c)_\BZ$ such that $\hat{F} := \Phi^{-1} \circ F|_{W^c} \circ \Phi$ and $\hat{G} := \Phi^{-1} \circ G|_{W^c} \circ \Phi$ satisfy the premises of Theorem \ref{mainresult2}, and that if $\Phi^{-1}(A')$ is Presburger in $\BZ^{d'}$, then $A'$ is Presburger in $\BZ^d$. 
\end{lemma}

\begin{proof}
Scaling a basis over $\BQ$ to a basis over $\BZ$, we have that $(W^c)_\BZ$ has dimension $d'$ and is free since it is a subgroup of $\BZ^d$. Choosing a basis $v_1, \ldots, v_{d'}$ for $(W^c)_\BZ$, let $\Phi: \BZ^{d'} \rightarrow (W^c)_\BZ$ be the corresponding isomorphism. Then $\hat{F}$ and $\hat{G}$ are both injective as they are compositions of injective functions, and  commute since $F$ and $G$ do. Further, $F((W^c)_\BZ) \subseteq F(\BZ^d) \cap F(W^c) \subseteq \BZ^d \cap W^c = (W^c)_\BZ$ since $F$ is integral and $W^c$ is $F$-invariant, so that $\hat{F} \in M_{d' \times d'}(\BZ)$. The same can be said for $\hat{G}$. The eigenvalues of $\hat{F}$ and $\hat{G}$ are subsets of those of $F$ and $G$ respectively, and so they are distinct, have common moduli $\lambda$ and $\mu$ respectively, and the ratios are never a root of unity. Also, by Lemma \ref{FequalsGinvariant}, $W^c$ is a sum of eigenlines of $F$, so the eigenvectors of $F$ lying in $W^c$ form a basis for $W^c$, with $F|_{W^c}$ and $G|_{W^c}$ simultaneously diagonalizable in it. Thus $\hat{F}$ and $\hat{G}$ are diagonalizable as well. We also have that
\begin{align*}
\Phi([w]_{\hat{F}}) &= \Phi\left(\sum\limits_{i=0}^n \hat{F}^i d_i\right) \\
&= \Phi\left(\sum\limits_{i=0}^n \Phi^{-1}F^i\Phi d_i\right) \\
&= \sum\limits_{i=0}^n F^i(\Phi d_i) \\
&= [\Phi w]_F
\end{align*}
whenever $w \in \Phi^{-1}(D_{F|_{W^c}})^*$. Hence given a $(D_{F|_{W^c}}, F|_{W^c})$-automaton for $A'$, we may produce a $(\Phi^{-1}(D_{F|_{W^c}}), \hat{F})$-automaton for $\Phi^{-1}(A')$. The same thing holds for $G$.

Since $\Phi$ is an integer matrix, suppose that $\Phi^{-1}(A')$ is Presburger definable with formula $\varphi(x)$. Then $A' = \{y \in \BZ^d: \exists x \in \BZ^{d'} (y = \Phi x \land \varphi(x))\}$ is Presburger definable in $\BZ^d$. 
\end{proof}

We are ready to prove our main result.

\begin{proof}[Proof of Theorem \ref{mainresult2}]
We proceed by induction on the dimension $d$. If $d=1$, then there are vacuously no non-trivial $F$-invariant subspaces. Thus we may apply Theorem \ref{mainresult1}, which gives us that $\chi_A$ is periodic on $(-\infty, a]$ and $[b, \infty)$, possibly with different periods, and is $1$ on finitely many singletons from $(a, b)$. Then $A \cap (-\infty, a]$ and $A \cap [b, \infty)$ are each a finite union of arithmetic progressions intersected with a ray, and $A \cap (a, b)$ is a finite set, hence $A$ is Presburger definable.

Now let $d \geq 2$ and suppose that the result holds in dimensions less than $d$. Let $M_F$ and $M_G$ be automata which compute $A$ in bases $(D_F, F)$ and $(D_G, G)$ respectively. Choose digit sets $D_F'$ and $D_G'$ in accordance with Lemma \ref{digitsetlemma}, and let $M_F' = (Q_F, D_F', \delta_F, q_{0, F}, S_F)$ and $M_G' = (Q_G, D_G', \delta_G, q_{0, G}, S_G)$ be the automata computing $A$ in bases $(D_F', F)$ and $(D_G', G)$ respectively given by Lemma \ref{expanddigitlemma}. In order to apply Lemma \ref{pumpspan}, we view $M_F'$ and $M_G'$ as DFAOs over the output alphabet $\{0, 1\}$, and replace them by their minimized versions in the sense of Myhill-Nerode. Let $Q_\infty$ be the set of $q \in Q_F$ such that $[L_{q, F}]_F$ is infinite. Let $s \in Q_\infty$, let $t$ be the state given by Lemma \ref{pumpspan} for $s$, and let $x \in [L_{s, F}]_F$. Let $x_1, y_1, \ldots, x_r, y_r \in [L_{s, F}]_F \cap [L_{t, G}]_G$ be the pairs given by Lemma \ref{pumpspan}, and define
\begin{align*}
U_s = \mathrm{span}_\BQ\{x_i - y_i\}
\end{align*} 
to be the subspace over $\BQ$ from Lemma \ref{pumpspan} corresponding to state $s$. This lemma establishes that $U_s$ is $F^l$-invariant for some $l \geq 1$, but by Lemma \ref{FequalsGinvariant} this implies that it is $F$ and $G$ invariant as well. Fix one such pair $(x_i, y_i)$ and let $x \in [L_{s, F}]_F$ and $z \in \BZ^d$. We have that
\begin{align*}
F^m x + z \in A &\iff F^m x_i + z \in A \text{ since $x, x_i$ both go to $s$ in the $F$-automaton} \\
&\iff G^n x_i + z + (F^m - G^n)x_i \in A \\
&\iff G^n y_i + z + (F^m - G^n)x_i \in A \text{ since $x_i, y_i$ both go to $t$ in the $G$-automaton} \\
&\iff F^m y_i + z + (F^m - G^n)(x_i - y_i) \in A \\
&\iff F^m x + z + (F^m - G^n)(x_i - y_i) \in A,
\end{align*}
provided we have that $z$ and $z + (F^m - G^n)(x_i - y_i)$ have an $F$-representation with at most $m$ digits and $z + (F^m - G^n)x_i$ has a $G$-representation with at most $n$ digits. Let $T$ be the maximum over $\|x_i - y_i\|_V$ and $\|x_i\|_V$ across all generators $x_i - y_i$ for $U_s$ across all $s \in Q_\infty$, and suppose that $\|z\|_V \leq (1-\delta)L\min\{\lambda^m, \mu^n\}$ for some $\delta < \frac{1}{2}$. By equivalence of norms, there exists a constant $C$ such that $\|F^m - G^n\|_\infty \leq C\|F^m - G^n\|_V$. Choose $\eta$ small and fix $\epsilon$ small enough so that $(1-\delta)L + \epsilon T < L$ and $|Q_\infty|TC\epsilon < \frac{1}{2} - \delta - \frac{\eta}{2}$. By Lemma \ref{closepowerslemma}, there exist $m$ and $n$ arbitrarily large such that
\begin{align*}
\|z + (F^m - G^n)(x_i - y_i)\|_V &\leq \|z\|_V + \|F^m - G^n\|_V\|x_i - y_i\|_V \\
&\leq (1-\delta)L\min\{\lambda^m, \mu^n\} + \epsilon \min\{\lambda^m, \mu^n\}T \\
& < L\min\{\lambda^m, \mu^n\},
\end{align*}
and the same holds for $z + (F^m - G^n)x_i$ as well. There exist finitely many bad points which do not lie in $[L_{s, F}]_F$ for any $s \in Q_\infty$, and so there exists $\alpha$ such that $\{y \in \BZ^d: \|y\|_\infty \geq \alpha\}$ avoids these points. Let $c_0 \in \BZ$ be large enough so that $\{y \in \BC^d: \|y\|_V \leq L\} \subseteq \{y \in \BC^d: \|y\|_\infty \leq c_0\}$. Let $P_m' := \{w \in \BZ^d: \|F^{-m}w\|_\infty < \alpha + c_0\}$.  

Note that since the image of $\BZ^d$ in $F^m$ is a lattice, the farthest point from any point in the image of $F^m$ lies in the center of the parallelipiped, and so the maximum possible distance this can be is
\begin{align*}
\frac{1}{2} \max_{\sigma \in \{\pm 1\}^d} \left\|F^m\left(x + \sum\limits_{j=1}^d \sigma_j e_j\right) - F^m(x)\right\|_V &= \frac{1}{2}\max_{\sigma \in \{\pm 1\}^d}\left\|F^m\left(\sum\limits_{j=1}^d \sigma_j e_j\right)\right\|_V \\
&\leq \frac{1}{2}\lambda^m \max_{\sigma \in \{\pm 1\}^d}\left\|\sum\limits_{j=1}^d \sigma_j e_j\right\|_V \\
&\leq \frac{1}{2}L\lambda^m,
\end{align*}
and $\frac{1}{2}L\lambda^m \leq (1-\delta)L\min\{\lambda^m, \mu^n\}$ by Lemma \ref{closepowerslemma}.

Suppose $w \not \in P_m'$. Analogous to the above, if $x \in \BZ^d$ is the nearest integer vector to $F^{-m}w$ in $V$-norm, we have that $\|F^{-m}w - x\|_V \leq \frac{1}{2}L$, and so $\|F^{-m}w - x\|_\infty \leq c_0$, so that $\|x\|_\infty \geq \|F^{-m}w\|_\infty - c_0 \geq \alpha$, so $x$ is not a bad point. If we write $w = F^m x + z$, then $\|z\|_V = \|F^m(F^{-m}w - x)\|_V \leq \frac{1}{2}L\lambda^m \leq \frac{1}{2}(1+\eta)L\min\{\lambda^m, \mu^n\}$. From this, for every $y$ outside of $P_m'$, there exists $U_s = \mathrm{span}_\BQ\{x_i - y_i\}$ for which we are periodic with respect to each $(F^m - G^n)(x_i - y_i)$ in a ball of radius
\begin{align*}
(1-\delta)L\min\{\lambda^m, \mu^n\} - \frac{1}{2}(1+\eta)L\min\{\lambda^m, \mu^n\} = (\frac{1}{2} - \delta - \frac{\eta}{2})L\min\{\lambda^m, \mu^n\}.
\end{align*}
By our choice of $\epsilon$, for sufficiently large $m$ and $n$ this ball certainly contains the ball of radius
\begin{align*}
k = |Q_\infty|TC\epsilon \min\{\lambda^m, \mu^n\} \leq (\frac{1}{2} - \delta - \frac{\eta}{2})L\min\{\lambda^m, \mu^n\}.
\end{align*}
These periods we have produced span $(F^m - G^n)U_s$. Since $U_s$ is $F$ and $G$-invariant, it is also $F^m$ and $G^n$-invariant, hence $(F^m - G^n)$-invariant, and $F^m - G^n$ is injective, so that $(F^m - G^n)U_s = U_s$. If the $x_i - y_i$ are not all linearly independent, pass to a subset of them which form a basis for $U_s$, so that their $(F^m - G^n)(x_i - y_i)$ counterparts form a periodic basis locally. To use Remark \ref{muchnikremark}, we require the estimate for $T_\mathcal{B}$ to be defined in terms of $\|.\|_\infty$ rather than $\|.\|_V$. Through scaling $\epsilon$ back by a constant, using equivalence of norms, this is sufficient in order to apply the generalized Muchnik's criterion. 

In order to apply Proposition \ref{muchnikcriterion}, it remains to show that every section $\Sec_{W, \ell, t_k}(A)$ is definable. By Lemma \ref{secinherit}, we have that $\Sec_{W, \ell, t_k}(A)$ is $F|_{W^c}$-automatic. By Lemma \ref{FequalsGinvariant}, since $W^c$ is $F$-invariant, it is $G$-invariant as well, and so by the same argument applied to $G$, $\Sec_{W, \ell, t_k}(A)$ is also $G|_{W^c}$-automatic. This set lies in $W^c$, and so it is of some dimension $d' < d$. If $d' = 0$, then it is $\emptyset$ or $\{0\}$, hence trivially definable. Otherwise, let $\Phi: \BZ^{d'} \rightarrow (W^c)_\BZ$ as in Lemma \ref{transferlemma}. By this lemma, it suffices to show that $\Phi^{-1}(\Sec_{W, \ell, t_k}(A))$ is Presburger definable over $\BZ^{d'}$. But it satisfies all the premises of the result, and so by the induction hypothesis this is true since $d' < d$. Hence the result is proven for dimension $d$ as well, and by induction it holds for all $d \geq 1$. 
\end{proof}

\section{Characterizing When a Presburger Definable Set is $(D_F, F)$-Automatic}

Under the traditional notion of $k$-automaticity, a Presburger definable set is $k$-automatic for every integer $k \geq 2$. In the setting of $(D_F, F)$-automatic sets, it is too optimistic to hope for a direct analogue of this result. Instead, we prove the following:

\begin{thm}\label{mainresult3}
Let $F$ satisfy the assumptions of Theorem \ref{mainresult2} which do not specify a $G$. Then a Presburger definable set $A$ is $(D_F, F)$-automatic for some digit set $D_F$ if and only if $A$ is a finite Boolean combination of sets of the form $v + V$ and $v + (V \cap \{x: \pm \ell(x) \geq 0\})$ with $V$ a finitely generated subgroup of $\BZ^d$, $\mathrm{span}_\BQ V$ $F$-invariant, and with $\ell$ the primitive integral left eigenvector of $F$ with real eigenvalue, unique up to sign, if such an $\ell$ exists.
\end{thm}

We begin by producing a partial result in one direction.

\begin{defn}
A set $S \subseteq \BZ^d$ is said to be linear if $S = v + \{\sum\limits_{i=1}^k c_iw_i: c_1, \ldots, c_k \in \BN\}$ with $v \in \BZ^d$ and each $w_i \in \BZ^d$. A set $A$ has a semilinear representation if $A$ can be written as a finite union of linear sets.
\end{defn}

It is well known that a set $A$ is Presburger definable if and only if it has a semilinear representation. \\

\begin{prop}\label{FinvariantPresburger}
If there exists a representation $A = \bigcup\limits_{i=1}^k (v_i + V_i)$ with each $V_i$ of the form $\{c_1u_1 + \cdots + c_mu_m: (c_1, \ldots, c_m) \in \BZ^m\}$ and $\mathrm{span}_\BQ V_i$ $F$-invariant for every $i$, then $A$ is $(D_F, F)$-automatic for some digit set $D_F$.
\end{prop}

\begin{proof}
Fix $u_1, \ldots, u_k \in \BZ^d$ and let $V_i = \{c_1u_1 + \cdots + c_mu_m: (c_1, \cdots, c_m) \in \BZ^m\}$. Let $W = \mathrm{span}_\BQ V_i$. We first show that if $W$ is $F$-invariant, then $W_\BZ$ is $(D_F, F)$-automatic for $D_F$ chosen according to Lemma \ref{digitsetlemma}. Since $W$ is $F$-invariant, we may choose an $F$-invariant complement $W^c$, so that $\BQ^d = W \oplus W^c$. We create a machine that recognizes $W_\BZ$ as follows: let $P$ be the projection map onto $W^c$, and let $N$ be a common denominator for all entries of $P$. Let $Q = NP$. Since $W$ and $W^c$ are both $F$-invariant, writing $x \in \BQ^d$ as $x = x_1 + x_2$ with $x_1 \in W$ and $x_2 \in W^c$, $PFx = P(Fx_1) + P(Fx_2) = Fx_2$ since $Fx_1 \in W$ as $W$ is $F$-invariant. On the other hand, $FPx = F(Px_1 + Px_2) = Fx_2$, so $P$ commutes with $F$ over $\BQ$, and hence so does $Q = NP$. We have that $x \in W_\BZ \iff x \in \BZ^d$ and $Qx = 0$. We proceed in a manner similar to Lemma \ref{expanddigitlemma}. Namely, when we first read $w_n$,
\begin{align*}
F^{n-1}(Qw_1) + \cdots + F(Qw_{n-1}) + Qw_n = 0 \iff F^{n-2}(Qw_1) + \cdots + Qw_{n-1} = -F^{-1}(Qw_n)
\end{align*}
since $F$ is injective. The left side is equivalent to $Q(F^{n-2}w_1 + \cdots + w_{n-1})$, which is integral since $Q$ is an integer matrix. So we transition to a special rejecting state if $F^{-1}(Qw_n)$ is not integral. Otherwise, transition to a state recording the value of $u = -F^{-1}(Qw_n)$. If we are in state $S_u$ with input $w_1 \cdots w_i$ left to be read, then
\begin{align*}
F^{n-1}(Qw_1) + \cdots + F(Qw_{i-1}) + Qw_i = u \iff F^{i-2}(Qw_1) + \cdots + Qw_{i-1} = F^{-1}(u - Qw_i),
\end{align*}
so reject if the right side is not integral, otherwise transition to a state corresponding to $u' = F^{-1}(u - Qw_i)$. After we have read all input, accept if we are in state $S_0$. Now by the same argument as in Lemma \ref{expanddigitlemma}, any vector $u$ lies in $B_V[C\lambda^{-1}/(1-\lambda^{-1})]$, where $C$ is large enough so that $Q(D_F) \subseteq B_V[C]$, and every $u$ is integral. Hence there are only finitely many states we can ever travel to. 

Next, we would like an automaton which recognizes that we are in $V_i$, not just $W_\BZ$. The former is of finite index in the latter, so that $V_i \oplus (W^c)_\BZ$ is of finite index in $W_\BZ \oplus (W^c)_\BZ$, and so $V_i \oplus (W^c)_\BZ$ is of finite index in $\BZ^d$. Let $N' = [\BZ^d: V_i \oplus (W^c)_\BZ]$, so that $N'\BZ^d \subseteq V_i \oplus (W^c)_\BZ$ and $F(N'\BZ^d) \subseteq N'\BZ^d$ as $F$ has integer entries. Then reading from left to right, we may keep track of $[w]_F \mod N'\BZ^d$ via $x \mapsto Fx + w_i \mod N'\BZ^d$, and accept if we end up with an equivalence class lying in $V_i \oplus (W^c)_\BZ$. Then a right-to-left version also exists, and running this in parallel with the automaton for $W_\BZ$, we can establish whether $[w]_F \in V_i \oplus (W^c)_\BZ$ and $Q([w]_F) = 0$, which together is equivalent to having $[w]_F \in V_i$. 

From here, one can create a machine which recognizes $v_i + V_i$: for an input $w = F^{n-1}w_1 + \cdots + Fw_{n-1} + w_n$, we test whether $w-v_i \in V_i$. To do this, take an expansion $v_i = F^{k-1}x_1 + \cdots + Fx_{k-1} + x_k$ and then up to depth $k$, transition as if we had read $w_{n-i} - v_{k-i}$ instead of $w_{n-i}$ (expanding the digit set for our automaton recognizing $V_i$ if needed), and after this proceed as normal, producing a subcopy of the automaton for $V_i$ on every branch. Since automaticity is closed under finite unions, any $A = \bigcup\limits_{i=1}^k (v_i + V_i)$ is automatic.
\end{proof}

Towards a correspondence in the other direction, we will need to introduce some terminology and notation and prove some technical lemmas. We place this notation in the following definition so as to provide references to it when necessary. We will use these definitions for the remainder of the section.

\begin{defn}\label{cellsetup}
We will say that a cone $\Gamma$ is realized at $x \in \BZ^d$ over a set $S$ if there exists finite $P \subset \BZ^d$ with $\Gamma$ being the cone generated by $P$ and $x + \BN P \subset S$. We let $R_S(x)$ denote the set of cones realized at $x$ over $S$.

It is well known that if $S$ is semilinear, then it may be written in the disjunctive normal form as a union of polyhedra intersected with modular congruences. Moreover, by taking the LCM over all moduli, we may write each congruence as being modulo the same constant. Write
\begin{align*}
    S = \bigcup\limits_{k=1}^K \mathcal{P}_k \cap (c_k + M\BZ^d),
\end{align*}
where each polyhedron $\mathcal{P}_k$ can be defined without loss of generality by relations of the form $z: \langle a_{k, l}, z\rangle \leq b_{k,l}$ for some constants $a_{k, l}, b_{k, l}$, which we may again take without loss of generality to be integers. We may also remove any pieces $\mathcal{P}_k \cap (c_k + M\BZ^d)$ which contain no lattice points from this decomposition without consequence. Given $x \in \BZ^d$, define $I(x) = \{k: c_k \equiv x \mod M\}$, and write 
\begin{align*}
C_k(x) := \{\sigma \in \BR^d: \langle a_{k,l}, \sigma\rangle \leq 0 \text{, with equality allowed only when } \langle a_{k,l}, x\rangle \leq b_{k,l}\}.
\end{align*}
Define 
\begin{align*}U(x) = \bigcup\limits_{k \in I(x)} C_k(x).
\end{align*}
Let $H$ be the collection of $a_{k,l}$ across all indices $k$ and $l$. Given a sign vector $s: H \rightarrow \{-1, 0, 1\}$, define the \textit{cell associated with $s$} to be
\begin{align*}
c_s = \{\sigma \in \BR^d: \mathrm{sgn}(\langle a, \sigma\rangle) = s(a) \text{ for all } a \in H\}.
\end{align*}
\end{defn}

With all of this, we note that since membership in $C_k(x)$ depends only on the sign vector, $C_k(x)$ is a union of cells, and so $U(x)$ is as well, a fact which we will repeatedly use in the arguments that follow.

\begin{lemma}\label{realizedconesimage}
Let $y = F^{-j}(x-u)$ with $x \equiv u \mod F^j(\BZ^d)$, and let $S_{u, j} = \{y \in \BZ^d: F^j y + u \in S\}$. Then $R_{S_{u, j}}(y) = F^{-j}(R_S(x))$.
\end{lemma}

\begin{proof}
If $\Gamma \in R_S(x)$, then $x + \BN P \subset S$ for some finite $P$ generating $\Gamma$. In particular, $x + \BN (|\det F^j| P) \subset S$ as well and $\Gamma$ is generated by $|\det F^j| P$ as well. We have that $\det(F^j)I = F^j \mathrm{adj}(F^j)$. Passing $P$ into both sides, we have that $|\det F^j|(P) = F^j(\mathrm{adj}(F^j(P))) \subseteq F^j(\BZ^d)$ since $\mathrm{adj}(F^j)$ has integer entries, and so $F^{-j}(|\det F^j|(P)) \subseteq \BZ^d$. Hence $y + \BN F^{-j}(|\det F^j|(P)) \subseteq \BZ^d$, and
\begin{align*}
u + F^j(y + \BN F^{-j}(|\det F^j|(P))) &= u + F^j(F^{-j}(x-u) + \BN F^{-j}(|\det F^j|(P))) \\
&= x + \BN|\det F^j|P \subset S,
\end{align*} 
so that $y + \BN F^{-j}(|\det F^j|(P)) \subseteq S_{u, j}$ and $F^{-j}(|\det F^j|(P))$ generates $F^{-j}(\Gamma)$. Conversely, if $\Gamma \in R_{S_{u, j}}(y)$, then $y + \BN P' \subset S_{u, j}$ for some finite $P'$ generating $\Gamma$, so that $S \supset u + F^j(y + \BN P') = u + F^j(F^{-j}(x-u) + \BN P') = x + \BN F^j(P')$, so that $F^j(\Gamma) \in R_S(x)$.
\end{proof} 

\begin{lemma}\label{realizedconesfinite}
Let $S$ be Presburger definable. Then $\Gamma \in R_S(x) \iff \Gamma \cap \BZ^d \subseteq U(x)$. Consequently, $R_S(x)$ takes on finitely many values across all $x \in S$.
\end{lemma}

\begin{proof}
Working with the setup of Definition \ref{cellsetup}, let $N > \max_{k, l} |b_{k, l} - \langle a_{k, l}, x \rangle|$. We have $x + Ny \in \mathcal{P}_k$ for a given $y \in \BZ^d$ if and only if
\begin{align*}
& \langle a_{k, l}, x + Ny\rangle \leq b_{k, l} \text{ for all } l \\ 
\iff & N\langle a_{k, l}, y\rangle \leq b_{k, l} - \langle a_{k, l}, x\rangle \text{ for all } l
\end{align*}
Now $\langle a_{k, l}, y\rangle \in \BZ$. If it is negative, the statement always holds, and if it is positive, it can never hold by our assumption on $N$. If it is zero, then it holds precisely when $\langle a_{k, l}, x\rangle \leq b_{k, l}$ for all $l$, and over all cases, this occurs precisely when $y \in C_k(x)$. Now let $\Gamma \in R_S(x)$. By Gordan's Lemma, $\Gamma \cap \BZ^d$ is finitely generated, say by $g_1, \ldots, g_s \in \BZ^d$. Since $\Gamma$ is generated by $P$, we may write $g_i = \sum_j c_{i,j}p_j$ with $c_{i, j} \in \BQ^+$. Clearing denominators, we get that $Kg_i \in \BN P$ for $1 \leq i \leq s$, so that $x + K(\Gamma \cap \BZ^d) \subseteq x + \BN P \subset S$. It is also clear that if $x + K(\Gamma \cap \BZ^d) \subset S$, then $P' = \{Kg_1, \ldots, Kg_s\}$ witnesses that $\Gamma \in R_S(x)$, so $\Gamma \in R_S(x) \iff \exists K \geq 1: x + K(\Gamma \cap \BZ^d) \subset S$, and this is true for any multiple of $K$ as well by the same logic. Choosing $N$ to be a sufficiently large multiple of both $K$ and the common modulus $M$, we have that 
\begin{align*}
\Gamma \in R_S(x) &\iff x + N(\Gamma \cap \BZ^d) \subset S \\
&\iff \text{ for every } y \in \Gamma \cap \BZ^d \text{, there exists } k \text{ such that } x + Ny \in \mathcal{P}_k \\
&\iff \Gamma \cap \BZ^d \subseteq U(x).
\end{align*}
But this last condition is dependent only on the signs of $\langle a_{k,l}, y\rangle$ for all $k$ and $l$, whether $x \in \mathcal{P}_k$, and the value of $x \mod M\BZ^d$. Since there are finitely many possible values for these parameters, it follows that $R_S(x)$ takes on finitely many values across all $x \in S$.
\end{proof}

Of course, this does not prove that any given $R_S(x)$ need be a finite set, which we would like in order to apply a pigeonhole-style argument. To bypass this, we will need a rather specialized notion of maximality.

\begin{defn}\label{goodpiece}
We will call a piece $x + \BN P \subseteq S$ \textit{good} if the relative interior of the cone generated by $P$ lies entirely within some single cell $c_0 \subseteq U(x)$ with $\mathrm{span}_\BQ P = \mathrm{span}_\BQ c_0$ and $\mathrm{span}_\BQ P$ is maximal among all $\mathrm{span}_\BQ c$ for cells $c$ satisfying $c \subseteq U(x)$ and $c_0 \subseteq \bar{c}$, and \textit{bad} if $\mathrm{span}_\BQ P$ is not maximal among these.
\end{defn}

\begin{lemma}\label{goodrep}
If a set $S$ is Presburger definable, it can be written as a finite union of good pieces and singletons.
\end{lemma}

\begin{proof}
Given a piece $\mathcal{P}_k \cap (c_k + M\BZ^d)$ of $S$, we may cover this by finitely many pieces of the form $y + \BN P_k$ with $y \in \BZ^d$ and $P_k \subseteq M\BZ^d$ some fixed finite set with the cone of $P_k$ being $\mathrm{rec}(\mathcal{P}_k)$. If we have $\mathrm{rec}(\mathcal{P}_k) = \{0\}$, then $P_k = \emptyset$, and so $y + \BN P_k = \{y\}$, a singleton, which we do not require to be a good piece. Suppose that $\mathrm{rec}(\mathcal{P}_k)$ is non-trivial. We have that $\overline{c_s}$ replaces strict inequality with non-strict inequality, so that conditions $\langle a_{k,l}, \sigma\rangle > 0$ become $\langle a_{k,l}, \sigma\rangle \geq 0$ and similar for $\langle a_{k,l}, \sigma\rangle < 0$. Let $\Gamma = \mathrm{Cone}(P_k)$. Let $\sigma \in \Gamma$ and let $\tau \in \mathrm{relint}(\Gamma)$ be generic, so that for any constraint not vanishing on $\mathrm{span}_\BQ \Gamma$, $\langle a, \tau\rangle \neq 0$. Then for sufficiently small $\epsilon > 0$, the sign vector $s$ of $\sigma + \epsilon\tau$ is constant. Then $\sigma \in \overline{c_s}$ and so $\overline{c_s} \cap \mathrm{relint} \Gamma$ contains a relatively open neighbourhood of $\sigma + \epsilon\tau$ in $\mathrm{relint} \Gamma$, so that $\dim(\overline{c_s} \cap \Gamma) = \dim \Gamma$. Let $\Gamma_s = \overline{c_s} \cap \Gamma$, so that $\Gamma = \bigcup\{\Gamma_s: \dim \Gamma_s = \dim \Gamma\}$. If $P_k = \{p_1, \cdots, p_m\}$, then $\{(n_1, \ldots, n_m) \in (\BR^+)^m: \sum n_ip_i \in \Gamma_s\}$ is a rational polyhedral cone, so the lattice points are a finitely generated semigroup by Gordan's Lemma. Then $\BN P_k \cap \Gamma_s$ is the image under the mapping $n_i \mapsto n_ip_i$, so we may write $\BN P_k \cap \Gamma_s = \BN P_{k, s}$ for some finite set $P_{k,s}$. From this same mapping, we see that $\mathrm{Cone}(P_{k,s}) = \Gamma_s$. So at the lattice level we have $y + \BN P_k = \bigcup\{y + \BN P_{k,s}: \dim \Gamma_s = \dim \Gamma\}$, and so $\dim P_{k, s} = \dim P_k$. So by replacing $\Gamma$ by the various $\Gamma_s$ of the same dimension which split it, we have $\mathrm{relint}\Gamma_s$ contained inside a single cell $c_s$, and so $c_s \subseteq U(y)$ since $c_s$ contains a lattice point of $\Gamma$, $\Gamma \in R_S(y)$ so that $\Gamma \cap \BZ^d \subseteq U(y)$ by Lemma \ref{realizedconesfinite}, and $U(y)$ is a union of cells. After this refinement, suppose that some piece $y + \BN P$ is bad. Then there exists $c \subseteq U(y)$ with $c_0 \subseteq \overline{c}$ and $\mathrm{span}_\BQ c \supsetneq \mathrm{span}_\BQ P$. Split $\BN P$ into the finitely many congruence classes $r + \BN MP$ with $r \in \BN P$, and take $N' > \max_{k,l}\max_r |b_{k,l} - \langle a_{k,l}, y + r\rangle|$ with $N'$ a multiple of $M$. Consider $y + r + \BN(MP \cup \{N'q_1, \ldots, N'q_n\})$ with each $q_i \in c$ and the $q_i$ a spanning set for $\mathrm{span}_\BQ c$.  

Clearly $y + r + \BN(MP \cup \{N'q_1, \ldots, N'q_n\})$ contains $y + r + \BN MP$. We have 
\begin{align*}
r \in \mathrm{Cone(P)} \subseteq \overline{c_0} \subseteq \overline{c} \subseteq \overline{C_{k'}(y)}
\end{align*}
where $c$ lies in $C_{k'}(y)$, and so $\langle a, r\rangle \leq 0$ for every constraint of $C_{k'}(y)$, hence $C_{k'}(y+r) \supseteq C_{k'}(y)$. By design, we have $y + r \equiv y \mod M$ and so $k' \in I(y+r)$. Choose a point $y + r + (\sum M\alpha_ip_i + N'\sum{\beta_jq_j})$ in our new piece, with the $q_j$ integral. If we have $\beta_j = 0$ for all $j$, then our point lies in $y + r + \BN M P$, hence in $S$. Otherwise, as argued before, we have that this lies in $C_{k'}(y+r)$ if and only if for all relevant constraints,
\begin{align*}
\langle a, y + r + (\sum M\alpha_ip_i + N'\sum{\beta_jq_j})\rangle \leq b \\
\iff N'\langle a, \sum{\beta_jq_j}\rangle \leq b - \langle a, y + r\rangle - M\langle a, \sum \alpha_ip_i\rangle.
\end{align*}
Now since $\Gamma = \mathrm{Cone}(P) \subseteq \overline{C_{k'}(y)}$, $\sum \alpha_ip_i \in \overline{C_{k'}(y)}$ and so $\langle a, \sum \alpha_ip_i\rangle \leq 0$, and $c \subseteq C_{k'}(y)$ so that $\langle a, \sum \beta_jq_j\rangle \leq 0$ as well. If we have strict inequality, since the $q_j$ are integral we have $\langle a, \sum \beta_jq_j\rangle \leq -1$, and so this is satisfied by how we chose $N'$. If instead $\langle a, \sum \beta_jq_j\rangle = 0$, then $\langle a, q_j\rangle = 0$ on any index for which $\beta_j > 0$. Since these signs are constant on cell $c$, $\langle a, \cdot\rangle$ is identically zero on $c$. Since $c \subseteq C_{k'}(y)$, equality can only hold if $\langle a, y\rangle \leq b$, while $\langle a, r\rangle \leq 0$, so $\langle a, y+r\rangle \leq b$. Hence $y + r + \BN(MP \cup \{N'q_1, \ldots, N'q_n\}) \subseteq S$. Letting $\Gamma' = \mathrm{Cone}(MP \cup \{N'q_1, \ldots, N'q_n\})$, we may write any point in $\mathrm{relint} \Gamma'$ as $t + t'$ with $t \in \mathrm{Cone}(P) \subseteq \overline{c_0} \subseteq \overline{c}$ and $t' \in c$ (generated by the $q_j$ terms). Since $c_0 \subseteq \overline{c}$, every equality constraint of $c$ is an equality constraint of $c_0$, so both terms vanish. On strict inequality constraints of $c$, $t$ can either match the constraint's sign or vanish, while $t'$ will match the sign, so $t + t'$ matches the sign. Thus $\mathrm{relint}(\Gamma') \subseteq c$. Also $\mathrm{span}_\BQ \Gamma' = \mathrm{span}_\BQ (P \cup \{q_1, \ldots, q_n\}) = \mathrm{span}_\BQ c$ since $\mathrm{span}_\BQ P \subseteq \mathrm{span}_\BQ \overline{c_0} \subseteq \mathrm{span}_\BQ c$ and $\mathrm{span}_\BQ\{q_1, \ldots, q_n\} = \mathrm{span}_\BQ c$. Thus our new cone satisfies all the assumptions on the old one, but is of greater dimension since $\mathrm{span}_\BQ c \supsetneq \mathrm{span}_\BQ P$. After repeating this enlarging argument at most $d$ times, we force the pieces to become good. 
\end{proof}

\begin{defn}
For $x \in S$ and $\delta = \BR^+ p$ a realized ray of $x$, let $\mathcal{W}_S(x, \delta)$ be the set of maximal elements of $\{\mathrm{span}_\BQ \Gamma: \Gamma \in R_S(x), \delta \subseteq \Gamma\}$.
\end{defn} 

\begin{lemma}\label{maxconeequiv}
Given $\delta = \BR^+ p$, we equivalently have that $\mathcal{W}_S(x, \delta)$ is the set of maximal elements of $\{\mathrm{span}_\BQ c: c \subseteq U(x), p \in \overline{c}\}$ for cells $c$. Moreover, if $x + \BN P \subseteq S$ is a good piece, then $\mathrm{span}_\BQ P \in \mathcal{W}_S(x, \delta)$ for any ray $\delta \subseteq \mathrm{relint}(\mathrm{Cone}(P))$.
\end{lemma}

\begin{proof}
Suppose $c \subseteq U(x)$ is a cell with $p \in \overline{c}$. Choose $q_1, \ldots, q_n \in c \cap \BZ^d$ spanning $\mathrm{span}_\BQ c$ and let $\Gamma = \mathrm{Cone}(p, q_1, \ldots, q_n)$. Since $p \in \overline{c}$, $p$ weakly satisfies any constraint on $\langle a, \cdot\rangle$ in the definition of $c$, while any $\sum \beta_j q_j$ strongly satisfies them, so that any $\alpha p + \sum \beta_j q_j \in \Gamma$ with at least one $\beta_j > 0$ strongly satisfies them, meaning they lie in $c \subseteq U(x)$. If the ray $\BR^+ p$ is realized over $x$, then $\alpha p \in U(x)$, which handles the case where all $\beta_j$ are zero. Thus $\Gamma \cap \BZ^d \subseteq U(x)$, and so $\Gamma \in R_S(x)$ by Lemma \ref{realizedconesfinite}, with $\mathrm{span}_\BQ \Gamma = \mathrm{span}_\BQ c$ since $p \in \overline{c} \subseteq \mathrm{span}_\BQ c$. 

Conversely, if $\Gamma \in R_S(x)$ and $p \in \Gamma$, let $\sigma \in \mathrm{relint}(\Gamma)$ be generic, so that for any constraint not vanishing on $\mathrm{span}_\BQ \Gamma$, $\langle a, \sigma\rangle \neq 0$. As before, for $\epsilon > 0$ sufficiently small, the sign vector $s$ of $p + \epsilon\sigma$ is constant. Let $c_s$ be the cell corresponding to this. Then $p$ will never reverse signs, only potentially send some to zero, so that $p \in \overline{c_s}$. Now $p \in \Gamma$ and $\sigma \in \mathrm{relint}\Gamma$ gives that $p + \epsilon\sigma \in \Gamma + \mathrm{relint}\Gamma = \mathrm{relint}\Gamma$, so $c_s \cap \mathrm{relint} \Gamma$ contains a relatively open neighbourhood of $p + \epsilon\sigma$ in $\mathrm{relint} \Gamma$, and hence a rational point, which can be scaled to an integer point. Now since $\Gamma \in R_S(x)$, by Lemma \ref{realizedconesfinite} we have that $\Gamma \cap \BZ^d \subseteq U(x)$. In particular $\mathrm{relint}\Gamma \subseteq U(x)$, and so $c_s \subseteq U(x)$ since cells either completely lie inside $U(x)$ or are disjoint, and $c_s$ contains a lattice point. Then $\mathrm{span}_\BQ(c_s \cap \mathrm{relint}\Gamma) = \mathrm{span}_\BQ \Gamma$ and hence $\mathrm{span}_\BQ \Gamma \subseteq \mathrm{span}_\BQ c_s$.

Combining the conclusions of the above two paragraphs, we see that maximal elements of $\{\mathrm{span}_\BQ \Gamma: \Gamma \in R_S(x), \delta \subseteq \Gamma\}$ are precisely maximal elements of $\{\mathrm{span}_\BQ c: c \subseteq U(x), p \in \overline{c}\}$.

Let $x + \BN P \subseteq S$ be a good piece, with cell $c_0$ as in Definition \ref{goodpiece}. We claim that $\mathrm{span}_\BQ P \in \mathcal{W}_S(x, \delta)$ for any ray $\delta \subseteq \mathrm{relint}(\mathrm{Cone}(P))$. Suppose not, then there exists a ray $\delta$ and $\Gamma \in R_S(x)$ with $\delta \subseteq \Gamma$ such that $ \mathrm{span}_\BQ P \subsetneq \mathrm{span}_\BQ \Gamma$. Since $\delta \subseteq \mathrm{relint}(\mathrm{Cone}(P))$, write $\delta = \BR^+p \subseteq c_0$ for some $p \in c_0$. Then there exists $c \subseteq U(x)$ with $p \in \overline{c}$ and $\mathrm{span}_\BQ \Gamma \subseteq \mathrm{span}_\BQ c$. Since $p \in c_0$ and $p \in \overline{c}$, if $s'$ and $s$ are the sign vectors for $c_0$ and $c$ respectively, $s'(a) \in \{0, s(a)\}$, and so we must have $c_0 \subseteq \overline{c}$. Since $x + \BN P$ is good, $\mathrm{span}_\BQ P = \mathrm{span}_\BQ c_0$, and so we must have $\mathrm{span}_\BQ c_0 \subsetneq \mathrm{span}_\BQ c$, contradicting maximality.
\end{proof}

\begin{prop}\label{finalsemilinear}
Let $A$ be a Presburger definable set, and let $F$ satisfy the assumptions of Theorem \ref{mainresult2} which do not specify a $G$. If $A$ is $(D_F, F)$-automatic for some digit set $D_F$, then there exists a semilinear representation $\bigcup\limits_{i=1}^k (v_i + V_i)$ with $\mathrm{span}_\BQ(V_i)$ $F$-invariant for every $i$. 
\end{prop}

\begin{proof}
Let $S_{u, j}$ be as in the statement of Lemma \ref{realizedconesimage}. Let $v, w \in D_F^*$. We have that $A_{[v]_F, |v|} = A_{[w]_F, |w|}$ if and only if there is no distinguishing extension for $v$ and $w$ in a right-to-left $(D_F, F)$-automaton for $A$, and so by Myhill-Nerode there are finitely many distinct sets $A_{[w]_F, |w|}$ as $w$ ranges over $D_F^*$. Write $E = \{A_{[w]_F, |w|}: w \in D_F^*\}$ for this finite set. 

Given $S$ semilinear, let $H_S$ and $U_S(x)$ be the $H$ and $U(x)$ from Definition \ref{cellsetup}. By Lemma \ref{maxconeequiv}, $\mathcal{W}_S(x, \delta)$ is characterized by which cells lie in $U_S(x)$ (which is determined by $x \mod M$ together with the inequalities $\langle a, x\rangle \leq b$ which hold), and which cells have $p$ in their closure, (of which there are $3^{|H_S|}$ possibilities). Thus $\bigcup\limits_{x, \delta} \mathcal{W}_S(x, \delta)$ is a finite set of subspaces.

Now if $\Gamma \in R_A(x)$, then $F^{-j}(\Gamma) \in R_{A_{u,j}}(y)$ by Lemma \ref{realizedconesimage}. Write $u = F^j z + [r]_F$ where $|r| = j$. Then we have
\begin{align*}
\Gamma' \in R_{A_{u,j}}(y) \iff & y + \BN P' \in A_{u, j} \\
\iff & u + F^j(y + \BN P') \in A \\
\iff & [r]_F + F^j z + F^j(y + \BN P') \in A \\
\iff & z + y + \BN P' \in A_{[r]_F, |r|} \\
\iff & \Gamma' \in R_{A_{[r]_F, |r|}}(y+z),
\end{align*}
and $y + z \in A_{[r]_F, |r|}$ since $[r]_F + F^j(y+z) = [r]_F + (x-u) + F^j z = x \in A$. Thus $F^{-j}(\Gamma) \in R_{A_{[w]_F, |w|}}(x)$ for some $w \in D_F^*$ and $x \in A$. Each $A_{[w]_F, |w|}$ is Presburger since it is the preimage of $A$ under an affine map with an integer matrix. Moreover, since the map $\Gamma \rightarrow F^{-j}(\Gamma)$ is an inclusion-preserving bijection, maximal elements correspond, and since $F^{-j}$ is injective and linear, we have that $\mathcal{W}_{A_{u,j}}(y, F^{-j}(\delta)) = F^{-j}(\mathcal{W}_A(x, \delta))$, and this is equal to $\mathcal{W}_{A_{[w]_F, |w|}}(y+z, F^{-j}(\delta))$ for some $w \in D_F^*$. Then since there are finitely many distinct $A_{[w]_F, |w|}$ and each of them is Presburger definable (semilinear), $\bigcup\limits_{w \in D_F^*}\bigcup\limits_{x, \delta} W_{A_{[w]_F, |w|}}(x, \delta)$ is a finite set of subspaces, and $F^{-j}(\mathrm{span}_\BQ \Gamma)$ lies in this set for all $j$ whenever $\Gamma$ is realized by some good $x + \BN P$. By pigeonhole, there exists $j_1 < j_2$ with $F^{-j_1}(\mathrm{span}_\BQ \Gamma) = F^{-j_2}(\mathrm{span}_\BQ \Gamma)$, so that $F^{j_2 - j_1}(\mathrm{span}_\BQ \Gamma) = \mathrm{span}_\BQ \Gamma$. Since this is a subspace over $\BQ$, by Lemma \ref{FequalsGinvariant}, $F(\mathrm{span}_\BQ \Gamma) = \mathrm{span}_\BQ \Gamma$ as well. In particular, by Lemma \ref{goodrep}, $A$ has a semilinear representation that is a union of finitely many good pieces and singletons, so we may apply this to each good piece, and invariance follows trivially for singletons.
\end{proof}

We would like to extend Proposition \ref{FinvariantPresburger} from $\BZ$-spans of generators to $\BN$-spans, or in the language of what we have just done, cones. In order to do this, we need to be capable of tracking the signs of the generating vectors. The main issue with this is that $F$ need not behave well with the sign of any particular generator, as only the $\BQ$-span of $V_i$ itself is $F$-invariant: generators may be mapped to one another or even negated. The following lemma repairs this under an additional assumption.

\begin{lemma}\label{covectorAutomaton}
If a set $A$ is a finite Boolean combination of sets of the form $v + V$ and $v + V \cap \{x: \pm \ell(x) \geq 0\}$ with $V$ a finitely generated subgroup of $\BZ^d$, $\mathrm{span}_\BQ V$ $F$-invariant, and $\ell$ an integral left eigenvector of $F$ with real eigenvalue, then $A$ is $(D_F, F)$-automatic for some digit set $D_F$. 
\end{lemma}

\begin{proof}
We proceed in a similar manner to the $d=1$ case of the backward direction of the proof of Theorem \ref{mainresult1}. The only thing not proven in Proposition \ref{FinvariantPresburger} is the case where we intersect with a half-space cut out by $\ell$. Suppose that $\ell$ has real eigenvalue $\pm \lambda$, with $|\lambda| > 1$ based on our assumptions on $F$. In particular, since $\ell$ is integral, this means that $\lambda \in \BZ$ since $F$ has integer entries. Suppose we have determined a digit set $D_F$ for which $V$ is $(D_F, F)$-automatic, and suppose that we have $|\ell(x)| \geq \frac{1}{\lambda - 1}\max\limits_{d \in D_F} |\ell(d)|$. Then
\begin{align*}
|\ell(Fx + d)| &= |\pm \lambda\ell(x) + \ell(d)| \\
&\geq \bigg||\pm \lambda\ell(x)| - |\ell(d)|\bigg| \\
&= \lambda|\ell(x)| - |\ell(d)| \text{ since } \lambda|\ell(x)| \geq \frac{\lambda}{\lambda - 1}\max\limits_{d \in D_F}|\ell(d)| \geq \max\limits_{d \in D_F}|\ell(d)| \\
&\geq |\ell(x)| + \bigg[(\lambda -1)|\ell(x)| -  \max\limits_{d \in D_F}|\ell(d)|\bigg] \\
&\geq |\ell(x)|,
\end{align*}
so that once we exceed $\frac{1}{\lambda - 1}\max\limits_{d \in D_F}|\ell(d)|$, we permanently exceed it. The only other detail is that if $\ell F = -|\lambda| \ell$, we must also keep track of the parity of input length to determine the sign of $\ell(x)$. This will always work since 
\begin{align*}
\lambda|\ell(x)| \geq \frac{\lambda}{\lambda - 1}\max\limits_{d \in D_F}|\ell(d)| > \max\limits_{d \in D_F}|\ell(d)|,
\end{align*}
since $\max\limits_{d \in D_F}|\ell(d)| > 0$, otherwise $\ell$ would vanish on every $F$-expansion, hence on all of $\BZ^d$. From this, we see that adding another digit will always swap the sign, and so we should keep track of the exact value $\ell(x)$ for input $x$ until we exceed our bound, at which point we only have to track signs if we get the negative eigenvalue. Note that such values $\ell(x)$ are always integers: $\ell$ is integral and $\lambda \in \BZ$, so this can be tracked with finitely many states. This gives an automaton for $\{x: \ell(x) \geq 0\}$, so we can get an automaton for $V \cap \{x: \ell(x) \geq 0\}$ by following the proof Proposition \ref{FinvariantPresburger} and then intersecting (after enlarging to a common digit set via \ref{digitsetlemma}), and then for $v + V \cap \{x: \ell(x) \geq 0\}$ by again appealing to the proof of Proposition \ref{FinvariantPresburger}. If we have a union of such automata, simply enlarge to a common digit set, and then since automaticity is closed under boolean combinations, the result follows. 
\end{proof}

\begin{lemma}\label{fulldimcellsclosure}
Fix a Disjunctive Normal Form setup for a set $S$. Let $x \in \BZ^d$. Then we have that
\begin{align*}
\bigcup\{\mathrm{rec}(\mathcal{P}_k): k \in I(x), \dim\mathrm{rec}(\mathcal{P}_k) = d\} = \overline{\mathrm{int}(U(x))} = \bigcup\{\bar{c}: c \subseteq U(x) \text{ a cell and } \dim c = d\}.
\end{align*}
\end{lemma}

\begin{proof}
Let $U_d(x) = \bigcup\{c: c \subseteq U(x) \text{ a cell and } \dim c = d\}$ be the union of all full dimensional cells. Since there are finitely many such cells, $\overline{U_d(x)} = \bigcup\{\bar{c}: c \subseteq U(x) \text{ a cell and } \dim c = d\}$. Every cell of maximal dimension will only ever have strict inequalities, so such cells are open. Then $U_d(x)$ is open, and so $U_d(x) \subseteq \mathrm{int}(U(x))$, and so $\overline{U_d(x)} \subseteq \overline{\mathrm{int}(U(x))}$. Full dimensional cells are precisely those with $s(a) \neq 0$ for all $a \in H$, since otherwise the cell lies within $\ker a$, ($a$ being viewed as a vector), which cannot be of full dimension. Let $U_{<d}(x)$ be the union of the lower dimensional cells. Because there are finitely many such cells, $U_{<d}(x)$ has empty interior since each of the cells do, and so $\mathrm{int}(U(x))\setminus U_{<d}(x)$ is dense in $\mathrm{int}(U(x))$. Also, we have $\mathrm{int}(U(x))\setminus U_{<d}(x) \subseteq U_d(x)$. Thus $\mathrm{int}(U(x)) \subseteq \overline{\mathrm{int}(U(x))\setminus U_{<d}(x)} \subseteq \overline{U_d(x)}$, and so $\overline{\mathrm{int}(U(x))} \subseteq \overline{U_d(x)}$. Hence $\overline{\mathrm{int}(U(x))} = \overline{U_d(x)}$.

Let $k \in I(x)$. Since $\mathcal{P}_k = \{\sigma: \langle a_{k,l}, \sigma\rangle \leq b_{k,l}\}$, if $\sigma \in \mathrm{rec}(\mathcal{P}_k)$, then $p + t\sigma \in \mathcal{P}_k$ for all $t > 0$ and any $p \in \mathcal{P}_k$, so that $\langle a_{k,l}, p\rangle + t\langle a_{k,l}, \sigma\rangle \leq b_{k,l}$ for all $t$, which forces $\langle a_{k,l}, \sigma\rangle \leq 0$. Thus $\mathrm{rec}(\mathcal{P}_k) \subseteq \{\sigma: \langle a_{k,l}, \sigma\rangle \leq 0 \,\forall l\}$, and the reverse inclusion is trivial since $\mathcal{P}_k$ is non-empty. This is defined by a union of closures of cells (not one cell, since cells have signs for all $\langle a_{k,l}, \sigma\rangle$, even for other values of $k$), and so a full dimension $\mathrm{rec}(\mathcal{P}_k)$ is a union of closures of full dimensional cells. If $c \subseteq U(x)$ is a cell of full dimension, then $c \subseteq C_k(x) \subseteq \mathrm{rec}(\mathcal{P}_k)$ for some $k \in I(x)$. Conversely, if $c$ is a cell of full dimension contained within $\mathrm{rec}(\mathcal{P}_k)$, then it satisfies $\langle a_{k,l}, \sigma\rangle \leq 0$ for all $l$, and since it is full dimension, must thus satisfy $\langle a_{k,l}, \sigma\rangle < 0$ for all $l$. Then $c \subseteq C_k(x) \subseteq U(x)$. Hence we have that
\begin{align*}  
\bigcup\{\mathrm{rec}(\mathcal{P}_k): k \in I(x), \dim \mathrm{rec}(\mathcal{P}_k) = d\} = \bigcup\{\bar{c}: c \subseteq U(x) \text{ a cell and } \dim c = d\}. 
\end{align*}
\end{proof}

\begin{lemma}\label{equivsubsets}
Let $S$ and $T$ be Presburger definable with $x \in S$ and $y \in T$. Suppose that $R_T(y) = F^{-j}(R_S(x))$. Then $U_T(y) = F^{-j}(U_S(x))$ as subsets of $\BR^d$.
\end{lemma}

\begin{proof}
The main difficulty is that $U_T(y)$ and $U_S(x)$ are associated with the DNF decompositions of different sets, and so they will not have the same cells. Let $H_S$ and $H_T$ be the collections of $a_{k,l}$ in polyhedra in the DNF decompositions of $S$ and $T$ respectively. We will call these the $H_S$ and $H_T$ decompositions respectively. Call a cell $c$ an $H_S$-cell if it corresponds to the $H_S$ decomposition. Then if $c_s$ is an $H_S$-cell,
\begin{align*}F^{-j}(c_s) &= \{F^{-j}\sigma: \mathrm{sgn}\langle a, \sigma\rangle = s(a) \forall a \in H_S\} \\
&= \{\sigma: \mathrm{sgn}\langle a, F^j\sigma\rangle = s(a) \forall a \in H_S\} \\
&= \{\sigma: \mathrm{sgn}\langle aF^j, \sigma\rangle = s(a) \forall a \in H_S\}, 
\end{align*}
which is an $H_SF^j$ cell, where $H_SF^j = \{aF^j: a \in H_S\}$. Each $aF^j$ is still integral because $F$ has integer entries and non-zero since $F$ is invertible. Thus $F^{-j}(U_S)$ is a union of $H_SF^j$-cells. \\

Now $U_S(x)$ and $U_T(y)$ both contain $0$ and are closed under positive scaling (since cells are). We have that 
\begin{align*}
z \in U_T(y) &\iff \BR^+z \in R_T(y) \text{ by Lemma \ref{realizedconesfinite}} \\
&\iff \BR^+F^j z = F^j(\BR^+ z) \in F^j(R_T(y)) = R_S(x) \text{ by assumption }\\
& \iff F^j z \in U_S(x) \text{ by Lemma \ref{realizedconesfinite}} \\
& \iff z \in F^{-j}(U_S(x))
\end{align*}
so we have correspondence on $\BZ^d$. Since both contain $0$ and are closed under positive scaling, we may scale any such lattice point $z$, and so we have correspondence on $\BQ^d$. It remains to extend this to $\BR^d$. Let $H = H_SF^j \cup H_T$, so that the $H$-cells are a refinement of both the $H_SF^j$-cells and $H_T$-cells, and so we may write both $U_T(y)$ and $F^{-j}(U_S(x))$ as a union of $H$-cells. We claim that every $H$-cell contains a rational point. Let $c_s$ be such an $H$-cell. Let $H_0 = \{a \in H: s(a) = 0\}$. Letting $L = \bigcap\limits_{a \in H_0} \ker a$, we have that $c_s = L \cap \{\sigma: \mathrm{sgn}\langle a, \sigma\rangle = s(a) \forall a \in H\setminus H_0\}$. Since the latter set is only determined by strict inequalities, we have that $c_s$ is open in $L$. Since $L$ is defined by rational parameters $a$, it has a basis of rational vectors, and hence $L \cap \BQ^d$ is dense in $L$. Then $c_s \cap (L \cap \BQ^d)$ must be non-empty, so $c_s$ contains a rational point, say $q_s$. Since $U_S(x)$ and $U_T(y)$ are both unions of $H$-cells, we have that $c_s \subseteq F^{-j}(U_S(x)) \iff q_s \in F^{-j}(U_S(x)) \iff q_s \in U_T(y) \iff c_s \subseteq U_T(y)$. Thus the two sets are composed of the same cells, and hence are equal on all of $\BR^d$. 
\end{proof}

\begin{lemma}\label{maxdimtrichotomy}
Let $A \subseteq \BZ^d$ be a Presburger definable set that is $(D_F, F)$-automatic for some digit set $D_F$. Fix $x \in A$, and write $K(U_A(x)) = \bigcup\{\bar{c}: c \subseteq U_A(x), \dim c = d\}$. Suppose that a covector $\ell$ is such that there exists $p \in \ker \ell$ and an open neighbourhood $N \ni p$ such that $K(U_A) \cap N = \{v: \ell(v) \geq 0\} \cap N$. Then $\ell$ is a positive multiple of the unique primitive integral left eigenvector $\pm \ell_0$ of $F$ with real eigenvalue. Consequently, $K(U_A)$ is either $\BR^d, \BR^d \cap \{v: \pm \ell_0(v) \geq 0\}$, or $\emptyset$. 
\end{lemma}

\begin{proof}
Let $\mathcal{N}(K(U_A(x)))$ be the set of primitive integral covectors $\ell$ for which there exists $p \in \ker \ell$ and an open neighbourhood $N \ni p$ such that $K \cap N = \{x: \ell(x) \geq 0\} \cap N$. By Lemma \ref{fulldimcellsclosure}, $K(U_A(x)) = \overline{\mathrm{int}(U_A)}$. Since $F^{-j}$ is a linear bijection, 
\begin{align*}F^{-j}(K(U_A)) \cap F^{-j}(N) &= F^{-j}(K \cap N) \\
&= F^{-j}(\{x: \ell(x) \geq 0\} \cap N) \\
&= F^{-j}(\{x: \ell(x) \geq 0\}) \cap F^{-j}(N) \\
&= \{x: (\ell \circ F^j)(x) \geq 0\} \cap F^{-j}(N), \\
\end{align*}
with $F^{-j}(N)$ an open neighbourhood of $F^{-j}(p)$. Hence $\mathcal{N}(F^{-j} K) = \{\mathrm{prim}(\ell \circ F^j): \ell \in \mathcal{N}(K)\}$, where $\mathrm{prim}(\ell \circ F^j)$ is $\ell \circ F^j$ scaled down appropriately to be primitive. Also, if $\ell \in \mathcal{N}(K)$ then $\ker \ell \cap \partial K$ has dimension $d-1$, so $\ker \ell$ lies within the boundary of one of the cones that union to form $K$. There are finitely many such subspaces, and hence finitely many $\ell \in \mathcal{N}(K)$, since we took such $\ell$ to be primitive.

Choose $u$ such that $x \equiv u \mod F^j(\BZ^d)$, and let $y = F^{-j}(x-u)$. By Lemma \ref{realizedconesimage}, we have that $R_{S_{u,j}}(y) = F^{-j}(R_S(x))$. Thus by Lemma \ref{equivsubsets} with $S = A$ and $T = A_{u,j}$, we have that $F^{-j}(U_A(x)) = U_{A_{u,j}}(y)$, and so 
\begin{align*}F^{-j}(K(U_A(x))) = F^{-j}(\overline{\mathrm{int}(U_A(x))}) = \overline{\mathrm{int}(F^{-j}(U_A(x)))} = \overline{\mathrm{int}(U_{A_{u,j}}(y))} = K(U_{A_{u,j}}(y)).\end{align*}

Write $u = F^j z + [r]_F$ with $|r| = j$. From the proof of Proposition \ref{finalsemilinear}, $R_{A_{u,j}}(y) = R_{A_{[r]_F, |r|}}(y+z)$. Then by Lemma \ref{equivsubsets} with $S = A_{u,j}$, $T = A_{[r]_F, |r|}$, and $j=0$, we have that $U_{A_{u,j}}(y) = U_{A_{[r]_F, |r|}}(y+z)$. Let $v, w \in D_F^*$. We have that $A_{[v]_F, |v|} = A_{[w]_F, |w|}$ if and only if there is no distinguishing extension for $v$ and $w$ in a right-to-left $(D_F, F)$-automaton for $A$, and so by Myhill-Nerode there are finitely many distinct sets $A_{[w]_F, |w|}$ as $w$ ranges over $D_F^*$. Then $\{U_{A_{u,j}}(y)\}_{j \geq 0}$ is finite, hence $\{K(U_{A_{u,j}}(y)\}_{j \geq 0}$ is as well. Also, for fixed $S$, any $U_S(x)$ takes on finitely many values across all $x \in S$ since $U_S(x)$ is a union of cells, and there are finitely many cells. Therefore $\bigcup_{j \geq 0} \mathcal{N}(F^{-j} K(U_A(x)))$ is a finite set of covectors. Hence if $\ell \in \mathcal{N}(K(U_A(x)))$, then each $\mathrm{prim}(\ell \circ F^j)$ lies in a finite set, and so by pigeonhole there exists $j_1, j_2$ such that $\mathrm{prim}(\ell \circ F^{j_1}) = \mathrm{prim}(\ell \circ F^{j_2})$. Then $\ell \circ F^{j_1} = c \ell \circ F^{j_2}$ for some $c \in \BQ^+$, and so $\ell \circ F^{j_1 - j_2} = c\ell$, so that $\ell$ is a left eigenvector of $F^{j_1 - j_2}$. By the argument of Lemma \ref{FequalsGinvariant}, $F^{j_1 - j_2}$ also has distinct eigenvalues and the eigenspaces of $F^{j_1 - j_2}$ and $F$ coincide, so that $\ell$ is also a left eigenvector of $F$, with eigenvalue $c_0$ with $|c_0| = \lambda$. Then $\ell F = c_0 \ell$, and so for the $i$th entry, $(\ell F)_i/\ell_i = c_0$. Since $\ell$ is rational, this means that $c_0 \in \BQ$, and hence $c_0 = \pm \lambda$. 

For convenience, write $K = K(U_A(x))$. Suppose that $q \in \partial K$. We cannot simply take a separating hyperplane containing $q$, since $K$ is a union of convex sets, not convex itself. Let $\epsilon > 0$, so that $B(q, \epsilon)$ contains $a$ in the interior of a dimension $d$ cell from $K$ and $b$ in the interior of a dimension $d$ cell from $K^c$. If $d=1$, the only possible boundary point is $q=0$, where $K$ is a closed ray, and so any normal covector has trivial kernel. If $d \geq 2$, we may remove any $(d-2)$-dimensional or lower pieces of $B(q, \epsilon)$: the removed set has codimension $2$, hence what remains is path connected. Since we removed such pieces, the path from $a$ to $b$ must pass through a wall of dimension exactly $d-1$ at the first point where the path leaves $K$. Near a point of the wall's relative interior, the only cells present other than the wall itself are the two adjacent cells of full dimension, and only one of these two cells lies in $K$. If $\ell'$ is the normal to this wall, then $\ell' \in \mathcal{N}(K)$, so that $\ker \ell' = \ker \ell_0$, and so that $B(q, \epsilon)$ contains a point from $\ker \ell_0$. Since $\epsilon$ was arbitrary, $q \in \ker \ell_0$. 

Thus we have $\partial K \subseteq \ker \ell_0$ if $\ell_0$ exists, or $\partial K = \emptyset$ if it does not. In the latter case, $K$ is open, and we already knew it was closed, so it is either $\BR^d$ or $\emptyset$. In the former case, let $H_{\ell_0}^+ = \{x: \ell_0(x) > 0\}$ and $H_{\ell_0}^- = \{x: \ell_0(x) < 0\}$. We have that $K \cap H_{\ell_0}^+$ is closed in $H_{\ell_0}^+$ since it is closed, and open in $H_{\ell_0}^+$ since $\partial K \subseteq \ker \ell_0$, so it is $H_{\ell_0}^+$ or $\emptyset$. Similarly, we have that $K \cap H_{\ell_0}^- = H_{\ell_0}^-$ or $\emptyset$. 
\begin{itemize}
\item If $K \cap H_{\ell_0}^+ = H_{\ell_0}^+$ and $K \cap H_{\ell_0}^- = H_{\ell_0}^-$, then $K \supseteq \overline{H_{\ell_0}^+ \cup H_{\ell_0}^-} = \BR^d$
\item If $K$ lies in $H_{\ell_0}^+$, then $H_{\ell_0}^+ \subseteq K \subseteq \overline{H_{\ell_0}^+}$, so that $K = \{x: \ell_0(x) \geq 0\}$, and similarly for $H_{\ell_0}^-$
\item If $K$ lies in neither, then $K \subseteq \ker \ell_0$, but $K$ is a union of pieces of full dimension, so we have $K = \emptyset$.
\end{itemize}
\end{proof}

With this, we have everything we need to prove the our main result in this section.

\begin{proof}[Proof of Theorem \ref{mainresult3}]
The backward direction is given by Lemma \ref{covectorAutomaton}. For the forward direction, we proceed by induction on the dimension $d$ of the space $\BR^d$ that we lie in. We will handle the base case $d=1$ as we proceed. Write $A$ in disjunctive normal form as  $A = \bigcup\limits_{k=1}^K \mathcal{P}_k \cap (c_k + M\BZ^d)$. Consider the set
\begin{align*}
S_r = \bigcup \{\mathrm{rec}(\mathcal{P}_k): k \in I(r), \dim(\mathrm{rec}(\mathcal{P}_k)) = d\}
\end{align*}
of full dimensional recession cones $\mathrm{rec}(\mathcal{P}_k)$ for which the corresponding $c_k$ satisfies $c_k \equiv r \mod M$. 

If $I(r) = \emptyset$, then $S_r = \emptyset$. If not, we may choose some $x \in A \cap (r + M\BZ^d)$. From Lemma \ref{fulldimcellsclosure} with $I(x) = I(r)$ as well as Lemma \ref{maxdimtrichotomy}, we either have $S_r = \emptyset, \BR^d$, or $\{x: \pm \ell(x) \geq 0\}$, where $\pm \ell$ is the unique real left eigenvector of $F$ (if one exists). Let $B = \bigcup\limits_{r < M} (r + M\BZ^d) \cap S_r \cap \BZ^d$. We claim that the maximum dimension of a recession cone realized anywhere in $A \Delta B$ is strictly less than $d$. Write $A \Delta B = (A\setminus B) \cup (B\setminus A)$. Any point in $A\setminus B$ either lies in some $\mathcal{P}_k$ with $\mathrm{rec}(\mathcal{P}_k) < d$, or in some $\mathcal{P}_k \cap (S_r \cap \BZ^d)^c$ with $\mathrm{rec}(\mathcal{P}_k) = d$. The first of these cases is clear. For the second case, if $S_r = \emptyset$ then $A\setminus B = A$ has no full dimensional recession cones. If $S_r = \BR^d$, then $A\setminus B = \emptyset$ and the claim holds vacuously. Otherwise, $S_r = \BR^d \cap \{x: \ell(x) \geq 0\}$. We have that $\mathrm{rec}(\mathcal{P}_k) \subseteq S_r$ by definition, and so $A\setminus B = \mathcal{P}_k \cap \{x: \ell(x) < 0\} \cap \BZ^d$ must have a recession cone which satisfies $\ell(x) = 0$ at all points, and hence is contained within $\ker \ell$, with $\dim(\ker \ell) \leq d-1$. Now looking at $B\setminus A$, within any given $r + M\BZ^d$, suppose that $k \in I(r)$. Given that the polyhedron $\mathcal{P}_k$ is defined by the inequalities $\langle a_{k,l}, x\rangle \leq b_{k,l}$, by inverting these inequalities over $\BZ^d$ as $\langle a_{k,l}, x\rangle \geq b_{k,l}+1$, we may write the complement of $\bigcup\limits_{k \in I(r)} \mathcal{P}_k$ as $\bigcup\limits_j \mathcal{Q}_j$ with each $\mathcal{Q}_j$ a polyhedron. 

We claim that any $\mathrm{rec}(\mathcal{Q}_j \cap S_r \cap \BZ^d)$ is of dimension strictly less than $d$. Suppose not. We have that $\mathrm{rec}(\mathcal{Q}_j \cap S_r \cap \BZ^d) \subseteq \mathrm{rec}(\mathcal{Q}_j) \cap \mathrm{rec}(S_r \cap \BZ^d) \subseteq \mathrm{rec}(S_r \cap \BZ^d) = S_r$, with the last equality coming from the fact that $S_r$ can only be $\emptyset, \BR^d$, or $\{x: \ell(x) \geq 0\}$. Since $\mathrm{rec}(\mathcal{Q}_j \cap S_r)$ is of full dimension, $\mathrm{int}(\mathrm{rec}(\mathcal{Q}_j \cap S_r)) \neq \emptyset$. Then we cannot have $\mathrm{int}(\mathrm{rec}(\mathcal{Q}_j \cap S_r)) \subseteq \bigcup_{k \in I(r)} \partial(\mathrm{rec}(\mathcal{P}_k))$, since applying $\mathrm{int}$ to both sides would give $\mathrm{int}(\mathrm{int}(\mathrm{rec}(\mathcal{Q}_j \cap S_r))) = \mathrm{int}(\mathrm{rec}(\mathcal{Q}_j \cap S_r)) \subseteq \emptyset$. Then some direction $u \in \mathrm{int}(\mathrm{rec}(\mathcal{Q}_j \cap S_r))$ must lie in $S_r\setminus \bigcup_{k \in I(r)} \partial(\mathrm{rec}(\mathcal{P}_k)) \subseteq \bigcup_{k \in I(r)} \mathrm{int}(\mathrm{rec}(\mathcal{P}_k))$, and hence in $\mathrm{int}(\mathrm{rec}(\mathcal{P}_k))$ for some $k \in I(r)$. Since interiors are open sets, we may take $u$ to be rational, and thus in $M\BZ^d$ after appropriately scaling. Choose a lattice point $q \in \mathcal{Q}_j \cap S_r \cap (r + M\BZ^d)$. Then for any $t \in \BZ^+$, $q + tu \in \mathcal{Q}_j \cap S_r \cap (r + M\BZ^d)$ as well since $u \in M\BZ^d$ and $u \in \mathrm{rec}(\mathcal{Q}_j \cap S_r)$. Fix $p \in \mathcal{P}_k$. Since $u \in \mathrm{int}(\mathrm{rec}(\mathcal{P}_k))$, there exists $N \in \BN$ such that for any $t > N$, $u + \frac{1}{t}(q - p) \in \mathrm{rec}(\mathcal{P}_k)$, so that $tu + (q-p) \in \mathrm{rec}(\mathcal{P}_k)$ for $t > N$. But then $q + tu = p + (q-p) + tu \in \mathcal{P}_k$, a contradiction since we had $q + tu \in \mathcal{Q}_j$ as well, and $\mathcal{Q}_j$ was chosen such that it was disjoint from $\mathcal{P}_k$. 

With the claim proven, any recession cone realized somewhere in $A \Delta B$ has dimension at most $d-1$. By Lemma \ref{covectorAutomaton}, $B$ is $F$-automatic, and hence so is $A \Delta B$. Then by Proposition \ref{finalsemilinear}, we may write $A \Delta B = \bigcup\limits_{i=1}^N (v_i + V_i)$ with $W_i := \mathrm{span}_\BQ(V_i)$ $F$-invariant for every $i$. 

Since $F$ has distinct eigenvalues, $W_i$ and $W_i^c$ are unique $F$-invariant complements of one another over $\BQ$. Fix coset representatives $t_1, \ldots, t_s$ for $[\BZ^d: (W_i)_\BZ \oplus (W_i^c)_\BZ]$ as per Definition \ref{sectiondef} and write $v_i = u_i + t_{k_i} + w'$ with $u_i \in (W_i)^c_\BZ$ and $w' \in (W_i)_\BZ$. Then $(A \Delta B) \cap (v_i + W_i) = (u_i + t_{k_i}) + \Sec_{W_i^c, u_i, t_{k_i}}(A \Delta B)$, and so $A \Delta B = \bigcup\limits_{i=1}^N \left((u_i + t_{k_i}) + \Sec_{W_i^c, u_i, t_{k_i}}(A \Delta B)\right)$. By Lemma \ref{secinherit}, each $\Sec_{W_i^c, u_i, t_{k_i}}(A \Delta B)$ is $F|_{W_i}$-automatic.

Now we have that each $\mathrm{Cone}(V_i)$ is realized at $v_i$ over $A \Delta B$ and lies within a finite union of polyhedra $R_j$ from outside of $S_r$, and hence with $\dim \mathrm{rec}(R_j) \leq d-1$ for all $j$. If $z \in V_i$, then so is each $nz$, and so by pigeonhole, some $R_j$ contains $v_i + nz$ for infinitely many $n$. But then $z \in \mathrm{rec}(R_j)$. Taking this over all $z \in V_i$, we have that $V_i \subseteq \bigcup_j\mathrm{rec}(R_j)$, and so writing $V_i = \BN P$, we have $\mathrm{Cone}(P) \subseteq \bigcup_j \mathrm{rec}(R_j)$. Thus $\dim V_i \leq d-1$ since $\dim\mathrm{rec}(R_j) \leq d-1$ for all $j$. Hence $\dim W_i \leq d-1$ as well, and so $\dim \Sec_{W_i^c, u_i, t_{k_i}}(A \Delta B) = d' < d$ as well. If $d=1$, there is nothing further to be done since we either get the empty set or a singleton for each $(u_i + t_{k_i}) + \Sec_{W_i^c, u_i, t_{k_i}}(A \Delta B)$. If $d' = 0$, then the section must be either $\emptyset$ or $\{0\}$, so that the corresponding piece is a singleton, hence allowed. Otherwise, suppose that the hypothesis holds for $d' < d$. By Lemma \ref{transferlemma} with $G$ omitted and $W = W_i^c$, we may reduce to showing that $\Phi^{-1}(\Sec_{W_i^c, u_i, t_{k_i}}(A \Delta B))$ is of the desired form in $\mathbb{Z}^{d'}$ since $\Phi: \BZ^{d'} \rightarrow (W_i)_\BZ$ is an isomorphism. Note that $\Sec_{W_i^c, u_i, t_{k_i}}(A \Delta B)$ is Presburger since it is the preimage under the affine map $w \mapsto u_i + t_{k_i} + w$, and that $\Phi^{-1}(S) = \{x \in \BZ^{d'}: \Phi(x) \in S\}$ is Presburger whenever $S$ is, since $\Phi$ is an integer matrix. 

There can be at most one left eigenvector of $\hat{F} = \Phi^{-1} \circ F|_{W_i} \circ \Phi$ with real eigenvalue, since the eigenvalues of $\hat{F}$ are a subset of those of $F$. Provided such a left eigenvector $\ell$ of $F$ with real eigenvalue exists, the corresponding left eigenvector for $\hat{F}$ is given by $\ell \circ \Phi$. If $\ell \circ \Phi = 0$, then the real eigenline is not in $W_i$, and so $\hat{F}$ does not have a real left eigenvector. Otherwise, $(\ell \circ \Phi)((\Phi^{-1}(x)) \geq 0 \iff \ell(x) \geq 0$, so that $\Phi(\{x \in \BZ^{d'}: (\ell \circ \Phi)(x) \geq 0\}) = \{x \in (W_i)_\BZ: \ell(x) \geq 0\}$. Also, since $\Phi$ is an isomorphism, and since $\Phi(\BZ^{d'}\setminus X) = (W_i)_\BZ \cap (\BZ^d\setminus \Phi(X))$ with $(W_i)_\BZ$ an allowed piece, it preserves boolean combinations. Finally, if $\mathrm{span}_\BQ W$ is $\hat{F}$-invariant, then $\mathrm{span}_\BQ \Phi(W)$ is $F$-invariant. But by the induction hypothesis, since $d' < d$, the statement holds for $\Phi^{-1}(\Sec_{W_i^c, u_i, t_{k_i}}(A \Delta B))$, and so it holds for $\Sec_{W_i^c, u_i, t_{k_i}}(A \Delta B)$ as well, and consequently for $(u_i + t_{k_i}) + \Sec_{W_i^c, u_i, t_{k_i}}(A \Delta B)$ as well, and finally for $A \Delta B$ as a whole. The statement holds for $B$ as well, and hence it holds for $A = (A \Delta B) \Delta B$.
\end{proof}

\begin{cor}
Under the hypotheses of Theorem \ref{mainresult2}, a set $A \subseteq \BZ^d$ is $(D_F, F)$-automatic and $(D_G, G)$-automatic for some digit sets $D_F$ and $D_G$ if and only if $A$ is a finite boolean combination of sets of the form $v + V$ and $v + (V \cap \{x: \ell(x) \geq 0\})$, with $V$ a finitely generated $\BZ^d$-subgroup with $\mathrm{span}_\BQ(V)$ $F$-invariant and $\ell$ the unique (up to sign) primitive integral left eigenvector of $F$ with real eigenvalue, if it exists.
\end{cor}

\begin{proof}
For the forward direction, Theorem \ref{mainresult2} gives that $A$ is Presburger definable, and so Theorem \ref{mainresult3} applied to $F$ gives the result. For the reverse direction, apply Lemma \ref{covectorAutomaton} for $F$. Note that $\mathrm{span}_\BQ(V)$ is $G$-invariant as well by Lemma \ref{FequalsGinvariant}, and that since $F$ has distinct eigenvalues, $G$ is diagonal in $F$'s eigenbasis, so that $\ell G = \gamma \ell$ for some $\gamma$. Since $\ell$ is rational, so is $\gamma$, and so we must have $\gamma = \pm \mu$. Then we may also apply Lemma \ref{covectorAutomaton} for $G$.
\end{proof}

\section{Examples and Applications}
We provide several examples of maps $F$ and $G$ which satisfy the required properties of Theorem \ref{mainresult1}. We obtain by corollary in the case $D=1$ a conjecture of Hansel and Safer from \cite{hanselsafer}. \\

\begin{lemma}\label{rationalarctan}
Let $D$ be a squarefree positive integer, and $p$ and $q$ positive integers with $(p, q) = 1$. Then $\arctan(p\sqrt{D}/q)$ is only a rational multiple of $\pi$ in the following cases:

\begin{itemize}
    \item $D = 1$ and $p = q = 1$
    \item $D = 3$, $p = 1$, and $q = 1$ or $3$.
\end{itemize}
\end{lemma}

\begin{proof}
If $\arctan(p\sqrt{D}/q) = a\pi/b$ for some $a, b \in \BZ$ with $(a, b) = 1$, then, equivalently, $p\sqrt{D}/q = \tan(a\pi/b)$ and so 
\begin{align*}
(q^2 + Dp^2)/q^2 = 1 + Dp^2/q^2 = 1 + \tan^2(a\pi/b) = 1/\cos^2(a\pi/b)
\end{align*}
Hence $\cos^2(a\pi/b) = q^2/(q^2 + Dp^2)$. Then
\begin{align*}
(2\cos(a\pi/b))^2 = 4\cos^2(a\pi/b) = 4q^2/(q^2 + Dp^2) 
\end{align*}

Let $\zeta_b^a = e^{ia\pi/b} = \cos(a\pi/b) + i\sin(a\pi/b)$. Then $(\zeta_b^a)^{-1} = \cos(a\pi/b) - i\sin(a\pi/b)$, so that $2\cos(a\pi/b) = \zeta_b^a + (\zeta_b^a)^{-1}$. By an identity of Lehmer, $z^{-\varphi(b)/2}\Phi_b(z)$ is a monic polynomial in $z + z^{-1}$, where $\Phi_b(z)$ is the $b$th cyclotomic polynomial. Since $(a, b) = 1$, $\zeta_b^a$ is a root of $\Phi_{2b}(z)$, and so $2\cos(a\pi/b)$ is a root of $z^{-\varphi(2b)/2}\Phi_{2b}(z)$, and hence is an algebraic integer. Then $(2\cos(a\pi/b))^2$ is an algebraic integer as well, but it is also rational from our equation above, and so it must be an integer. Hence we must have
\begin{align*}
(q^2 + Dp^2) \mid 4q^2
\end{align*}
Let $m = (q, D)$ and write $q = um, D = vm$ with $(u, v) = 1$. We then have that $(u^2m^2 + vmp^2) \mid 4u^2m^2$ and so $(u^2m + vp^2) \mid 4u^2m$. We have that $(p, u^2m^2) = (p, q) = 1$, so $(p^2, u^2m) = 1$. Also, we must have $(v, m) = 1$ since $D = vm$ is squarefree. From this along with the fact that $(u, v) = 1$, we obtain that $(vp^2, u^2m) = 1$, and so $(vp^2 + u^2m, u^2m) = 1$. Then we must have $vp^2 + u^2m \mid 4$.

Then we either have $vp^2 + u^2m = 1, 2$, or $4$.

\begin{itemize}
\item If $vp^2 = 4$ then $u^2m = 0$ and so $q = 0$, hence this is inadmissible.
\item If $vp^2 = 3$ then $u^2m = 1$ and so $u = m = 1$ (hence $q = 1$). The only way this can happen is if $v = 3$ and $p = 1$. Then we get $D = 3, p = 1, q = 1$.
\item If $vp^2 = 2$ then $u^2m = 2$ (since from above $u^2m = 0$ is impossible), and so $u = 1$ and $m = 2$ (hence $q = 2$). This can only occur if $v = 2$ and $p = 1$, but then $D = vm = 4$, a contradiction since $D$ is squarefree.
\item If $vp^2 = 1$ then either $u^2m = 1$ or $u^2m = 3$. If the former, then $u = m = 1$ (hence $q = 1$). If the latter, then $u = 1$ and $m = 3$ (hence $q = 3$). Either way, this can only occur if $v = p = 1$. Then in the former case we have $D = 1, p = 1, q = 1$, while in the latter we have $D = 3, p = 1, q = 3$.
\end{itemize}

Among the possible solutions, if $D = p = q = 1$ then $\arctan(1) = \pi/4$, if $D = 3$ and $p = q = 1$ then $\arctan(\sqrt{3}) = \pi/3$, and if $D = 3, p = 1$, and $q = 3$ then $\arctan(1/\sqrt{3}) = \pi/6$, and so all three possibilities are realised.
\end{proof}

Using the above lemma, we show that Theorem \ref{mainresult1} applies to a wide class of imaginary quadratic integers. \\

\begin{prop}\label{imaginaryquadratic}
Let $D$ be a squarefree positive integer. Let $w, z \neq \pm 1 \in \BZ[\sqrt{-D}]$ be such that $|w|$ and $|z|$ are multiplicatively independent and non-zero, and suppose that a sequence $(a_n)_{n \in \BZ[\sqrt{-D}]}$ is both $(D_w, w)$ and $(D_z, z)$-automatic for digit sets $D_w, D_z \subseteq \BZ[\sqrt{-D}]$. Suppose further that:
\begin{itemize}
\item If $D = 1$, $w$ and $z$ are not both from the set
\begin{align*}\{q + p\sqrt{D}i: p = 0 \text{ or } q = 0 \text { or } p = \pm q\};\end{align*}
\item If $D = 3$, $w$ and $z$ are not both from the set
\begin{align*}\{q + p\sqrt{D}i: p = 0 \text{ or } q = 0 \text { or } q = \pm p \text{ or } q = \pm 3p\};\end{align*}
\item Else, $w$ and $z$ are not both from the set $\{q + p\sqrt{D}i: p = 0 \text{ or } q = 0\}$,
\end{itemize}
Finally, if $D = 1$, suppose that $w, z \neq \pm i$.

Then $(a_n)_{n \in \BZ[\sqrt{-D}]}$ is ultimately periodic.
\end{prop}

\begin{proof}
Note that we may represent a quadratic integer $x + y\sqrt{D}i$ as a vector in $\BZ^2$. Then multiplication by such a quadratic integer can be viewed as multiplication by the matrix $\begin{pmatrix}x & -Dy \\ y & x\end{pmatrix}$. So if $w = a + b\sqrt{D}i$ and $z = u + v\sqrt{D}i$, then set $F = \begin{pmatrix}a & -Db \\ b & a\end{pmatrix}$ and $G = \begin{pmatrix}u & -Dv \\ v & u\end{pmatrix}$. These have eigenvalues $a \pm b\sqrt{D}i$ and $u \pm v\sqrt{D}i$ respectively, and so clearly these eigenvalue pairs are of the same modulus, and the moduli of $F$ and $G$ are multiplicatively independent by assumption. Note that $|a + b\sqrt{D}i| = \sqrt{a^2 + Db^2}$. If $b = 0$ then $|a| \geq 2$ so that $\sqrt{a^2 + Db^2} > 1$. If $a = 0$, then $\sqrt{a^2 + Db^2} = |b|\sqrt{D}$. If $D > 1$ then $|b| \geq 1$ and so $|b|\sqrt{D} > 1$. If $D = 1$, we cannot be equal to $\pm i$, and so $|b| \geq 2$, so $|b|\sqrt{D} > 1$. If $a, b \neq 0$, then $\sqrt{a^2 + Db^2} \geq \sqrt{1 + D} \geq \sqrt{2} > 1$. In any case, our eigenvalues have modulus strictly greater than $1$. It is also clear that $F$ and $G$ commute since they inherit this from the quadratic integers they represent. They are also diagonalizable as they have distinct eigenvalues. Hence they are simultaneously diagonalizable. Note that any non-trivial invariant subspace of $F^m$ or $G^n$ for some positive integers $m, n$ must have dimension exactly $1$, since the underlying space $\BQ^2$ has dimension $2$. Such subspaces are precisely lines through the origin in $\BC$. Note that any imaginary quadratic integer acts on such a line via rotation by its argument. For $F^m$ to have a non-trivial invariant subspace, the corresponding quadratic integer $(a + b\sqrt{D}i)^m$ would have to have argument $0$ or $\pi$, making it real. This implies that $a + b\sqrt{D}i$ must have an argument that is a rational multiple of $\pi$. Specifically, for $a \neq 0$ we have that

\begin{align*}
\mathrm{Arg}(a + b\sqrt{D}i) = 
\begin{cases}
        \arctan(b\sqrt{D}/a) & \text{if } a > 0 \\
        \arctan(b\sqrt{D}/a) + \pi & \text{if } a < 0, b \geq 0 \\
        \arctan(b\sqrt{D}/a) - \pi & \text{if } a < 0, b < 0 \\
    \end{cases}
\end{align*}

If $a = 0$ then the argument is $\pi/2$ if $b > 0$ and $-\pi/2$ if $b < 0$. Either way, for $a \neq 0$ it is clear that the argument is a rational multiple of $\pi$ if and only if $\arctan(b\sqrt{D}/a)$ is. We have chosen our hypotheses specifically to avoid purely real and purely imaginary values, as well as values from Lemma \ref{rationalarctan} which produce values of arctan which are a rational multiple of $\pi$.

With this, all the requirements of Theorem \ref{mainresult1} are satisfied with dimension $d = 2$, and so the result follows.
\end{proof}

We remark that it is possible for two Gaussian integers to be multiplicatively independent but not have multiplicatively independent moduli. For example, $2+i$ and $2-i$ are non-associate Gaussian primes, while both have modulus $\sqrt{5}$. The result of Bustos-Gajardo, Fokkink, and Yassawi in \cite{Gaussian26} handles such cases, whereas our result does not. In spite of this, we can still prove Hansel and Safer's conjecture by corollary.

\begin{cor}\label{hanselsafer}
Let $p$ and $q$ be distinct positive integers, and let $D_p = \{0, 1, \ldots, p^2\}$ and $D_q = \{0, 1, \ldots, q^2\}$. Then a set $S \subseteq \BZ[i]$ is both $(D_p, -p+i)$ and $(D_q, -q+i)$-automatic if and only if there exists a periodic set $P$ such that $S \Delta P$ is finite.  
\end{cor}

\begin{proof}
The only direction not known to the authors of \cite{hanselsafer} is the forward direction. Recall that a set $S$ is $F$-automatic if and only if its characteristic sequence $(\chi_{S, n})_{n \in \BZ[i]}$ is $F$-automatic. Then Proposition \ref{imaginaryquadratic} applies in the case $D = 1$, since $p$ and $q$ are distinct positive integers, and so we cannot have the real or imaginary parts being zero, and at most one of $p$ and $q$ can be $1$, which is the only way for $-p + i$ to have real part equal to plus or minus its imaginary part. Moreover, $|-p+i| = \sqrt{p^2 + 1}$ and $|-q+i| = \sqrt{q^2 + 1}$. If these are multiplicatively dependent, then so are $p^2 + 1$ and $q^2 + 1$, and so since these are positive integers, they must both be powers of some common $b \in \BN$. By Lebesgue's Theorem, the equation $x^2 + 1 = y^k$ has no positive integer solutions for $k \geq 2$, so we would have to have $p^2+1=b=q^2+1$, hence $p = q$, a contradiction. Thus there is a periodic binary sequence $(p_n)_{n \in \BZ[i]}$ such that $(\chi_{S, n})$ and $(p_n)$ differ by finitely many terms. Let $P$ be the set whose characteristic sequence is $(p_n)$. Then $S \Delta P$ is finite.
\end{proof}

\newpage

\printbibliography

\end{document}